\documentclass[letterpaper]{JHEP3}
\usepackage{cite}
\usepackage{graphicx}
\usepackage{amsmath}

\newcommand{\be}{\begin{equation}}
\newcommand{\ee}{\end{equation}}
\newcommand{\bea}{\begin{eqnarray}}
\newcommand{\eea}{\end{eqnarray}}

\renewcommand{\b}[1]{\bar{#1}}
\renewcommand{\t}[1]{\tilde{#1}}
\newcommand{\Del}{\nabla}
\newcommand{\del}{\partial}

\newcommand{\comment}[1]{}
\newcommand{\tg}{\tilde{g}}

\newcommand{\beq}{\begin{equation} }
\newcommand{\eeq}{\end{equation}}

\newcommand{\lb}{\label}

\newcommand{\im}{\textnormal{Im}}
\renewcommand{\ap}{\alpha'}
\newcommand{\ket}[1]{|#1\rangle}
\newcommand{\bra}[1]{\langle #1|}

\title{Warped Spectroscopy: Localization of Frozen Bulk Modes}
\author{Andrew R. Frey\\ California Institute of Technology\\
Mail Stop 452-48\\ Pasadena, CA 91125, USA\\
\email{frey@theory.caltech.edu}}
\author{Anshuman Maharana\\ Department of Physics\\
University of California, Santa Barbara\\ Santa Barbara, CA 93106, USA\\
\email{anshuman@physics.ucsb.edu}} \preprint{hep-th/0603233 \\
CALT-68-2591}

\abstract{We study the 10D equation of motion of dilaton-axion fluctuations 
in type IIB string compactifications with three-form flux, taking
warping into account.  Using simplified models with physics comparable to
actual compactifications, we argue that 
the lightest mode localizes in long warped throats and takes a mass
of order the warped string scale.  Also, the 
Gukov-Vafa-Witten superpotential is
valid for the lightest mass mode; however, the mass is similar
to the Kaluza-Klein scale, so the dilaton-axion should be integrated
out of the effective theory in this long throat regime (leaving a 
constant superpotential).  
On the other hand, there is a large
hierarchy between flux-induced and KK mass scales for moderate or weak 
warping.  This hierarchy agrees with
arguments given for trivial warping.  Along the way, we also estimate
the effect of the other 10D supergravity equations of motion on the 
dilaton-axion fluctuation, since these equations act as constraints.  
We argue that they give negligible corrections to the simplest 
approximation.}


\begin{document}

\section{Introduction}\label{s:intro}

       A central theme in string phenomenology in recent years
has been the study of compactifications in the presence of
background fluxes. In these constructions, non-trivial quanta of
form fields wrap cycles of the internal manifold, and ``geometric flux''
may introduce nontrivial fibration structures.
One of the attractive features of these constructions is that they
help resolve the moduli problem in Calabi-Yau compactifications. 
With the introduction of flux, a large number of flat directions
of the Calabi-Yau moduli space 
are lifted classically. As a result, the moduli acquire specific expectation
values.  With the loss of flat directions, the number of massless
four dimensional scalars are reduced, helping make string compactifications
more realistic.

In type IIB string theory, the work of
Giddings, Kachru, and Polchinski \cite{hep-th/0105097} provides an
algorithm to construct explicit solutions to equations of motion
in the presence of fluxes . In these constructions, the geometry
is a warped product with a conformal Calabi-Yau as the internal
manifold. Imaginary-self-dual three-form flux threads the cycles of
the Calabi-Yau, and these fluxes generically freeze the complex
structure moduli of the Calabi-Yau as well as the dilaton-axion. 
Detailed studies of this
phenomenon have been carried out in a large number 
of examples \cite{hep-th/0201028,hep-th/0201029,hep-th/0301139,hep-th/0312104,
hep-th/0505260,hep-th/0506090}.  For reviews, see
\cite{hep-th/0308156,hep-th/0405068,hep-th/0509003}.

       The four dimensional effective action governing the low
energy dynamics of these compactifications is of much interest for
a large number of phenomenological questions. 
However, at present we do not
have a complete understanding of the dimensional
reduction of these backgrounds; the warp factor gives rise to many
complications  \cite{hep-th/0307084,hep-th/0312076,hep-th/0407126}. 
Steps toward understanding the dimensional reduction were
taken in \cite{hep-th/0507158}, in which the equations describing
linearized fluctuations were derived. A complete analysis was not
possible because of two reasons. First, the equations are
extremely complicated in the presence of regions of large warping.
Second, the problem has the nature of an eigenvalue problem; that is,
a solution involves finding both the mass and the wavefunction of
the excitation.

Let us examine these issues in more detail.  
In the infinite volume limit, the warping is
small in all regions of the spacetime and the linearized equations are
easily analyzed. 
We find fluctuations of the complex structure moduli and
dilaton-axion of mass suppressed by $\ap/d^2$ compared to the Kaluza-Klein
scale, where $d$ is the linear scale of the compactification.
The Kaluza-Klein wavefunctions of the metric perturbations are just the 
complex structure deformations of the Calabi-Yau, while the
wavefunction of the dilaton-axion is a constant over the compactification.

However, at smaller volume, the compactification
develops throats, regions of significant warping --- the warp factor
becomes nontrivial.  Because the warp factor has nonvanishing derivatives
along the compact directions, the linearized supergravity (SUGRA)
equations of motion couple, so the fields in the four dimensional 
effective theory are actually mixtures of the SUGRA fluctuations.  The 
extra SUGRA fields that are excited in a particular 4D modulus are called
``compensators'' \cite{hep-th/0507158}, as they compensate for 
constraints imposed by the coupled equations.  The warp factor also 
introduces position dependence to the fluctuation equations.  Therefore,
the true Kaluza-Klein modes (denoted ``KK'' modes) are actually 
mixtures of the naive Kaluza-Klein (``kk'') modes of the unwarped
Calabi-Yau.  

        Given the interesting phenomenological applications of
regions of large warping, it is important to understand 
Kaluza-Klein reduction in their presence. In this paper, we 
take some preliminary steps in this direction. Our focus
is the dilaton-axion, which has the simplest equations of
motion.  To make the equations of
motion tractable, we introduce some simplifications into the equations
of motion and the geometry. 
While in agreement with our understanding of the
infinite volume limit, our models suggest a qualitatively distinct
behavior in the presence of deep throats. In particular, when the warped
hierarchy between the top and bottom of the throat is large, 
the lowest mass mode of the dilaton has a mass given by the warped string
scale,
and its wavefunction is highly localized in the bottom of the
throat. This mode is continuously connected to the lowest mode at
infinite volume --- as the volume of the compactifaction is
decreased, the wavefunction of the dilaton continuously varies from
being uniformly spread over the internal manifold to a highly
localized function.  The fact that the lowest mode changes continuously as
a function of the compactification volume modulus also implies that
the Gukov-Vafa-Witten superpotential \cite{hep-th/9906070} 
applies to the lightest KK mode even when warping is important.  We also 
find that the lightest mass is generically similar to the KK mass gap 
in the long throat regime,
so the dilaton-axion should often be integrated out of the effective
theory, leaving a constant superpotential.

Our conclusions are similar to those of Goldberger and Wise
\cite{hep-ph/9907218}, who discussed the dimensional reduction
of a massive scalar in the Randall-Sundrum scenario
\cite{hep-ph/9905221}. In fact, our estimate of the mass in 
the long throat regime reproduces that of \cite{hep-ph/9907218}.
In the context of type IIB string compactifications,
the Kaluza-Klein modes of the graviton exhibit a similar
phenomenon \cite{hep-th/0512076,hep-th/0512249,hep-th/0602136},
except that the lowest mode is massless and does not localize.
Although we are unable to perform a systematic study, our
analysis suggests that localization in long throats 
also occurs for the complex structure moduli of the underlying 
Calabi-Yau.

In outline, we next present a toy model which exhibits most
of the key features of our analysis. Section \ref{s:geometry} 
reviews some
aspects of flux compactifications in type IIB string theory. 
The bulk of the paper is section \ref{s:eom}, in which we first 
discuss the equations of motion for SUGRA fluctuations and argue that
the compensators give negligible corrections to the mass.  Then we
introduce our simplified geometry in \ref{s:simpleeom}, reproducing
the results of \cite{hep-ph/9907218} and give a 
semiclassical analysis in \ref{s:semiclassical}.  We then give results
for compactification-like models in section \ref{s:detailed}.
Finally, in section \ref{s:discuss}, we discuss various implications of
our results and some future directions.

 \subsection{Toy Example}\lb{s:toy}

Consider the five dimensional spacetime
\begin{equation}
ds^{2} = e^{2A(\phi)} \eta_{\mu \nu} dx^{\mu} dx^{\nu} +
d \phi^{2}  \  \  \ 0 \leq \phi \leq R \lb{met}
\end{equation}
where the warping is a step function
 \begin{eqnarray}
  \nonumber A(\phi) &=& -H\ , \ \  0 \leq \phi \leq L  \ \ \
(  \textnormal{region I}  ) \nonumber \\
A(\phi) &=& 0\ , \ \  \ L <  \phi \leq  R  \ \ \  (
\textnormal{region II} )\ . \lb{stwp}
\end{eqnarray}
      In this simplified model, region II
is the analogue of the bulk in a string
construction with fluxes, the part of the Calabi-Yau where the warping is
insignificant. Region I represents a region of
non-trivial warping; for $ H \gg 1$ and $ R \gg L $, it can be thought
of as a throat in the compactification.

         We consider a scalar particle $\tau$, of five-dimensional (5D)
mass $m_{f}$ (meaning ``flux-induced mass,'' as will be clear later),
propagating in this spacetime with
Neumann boundary conditions.  We dimensionally reduce this
scalar along the $\phi$ direction, particularly examining the
wavefunction and masses of particle-like excitations in 4D
as a function of $H$ and $m_{f}$.

The equation of motion for the scalar  is
\begin{equation}
  \left(\partial_{\phi} e^{4A(\phi)} \partial_{\phi}  +
  e^{2A(\phi)}\partial^{\mu}  \partial_{\mu} \right) \tau( \phi, x)
 = e^{4A(\phi)}m_{f}^{2}  \tau( \phi, x)\ . \label{de}
\end{equation}
  For 4D particle excitations, the field decomposes as $ \tau( \phi, x) =
  \tau_{n} ( \phi ) u_{n}( x ) $, with $ \partial^{\mu}
  \partial_{\mu}  u_{n}( x ) = m^{2}_{n} u_{n}( x ) $. Equation
  (\ref{de}) then becomes
\begin{equation}
 \frac{d }{ d \phi} \left( e^{4A(\phi)}
\frac{ d \tau_{n}}  {d \phi } \right)     + m_{n}^{2}e^{2A(\phi)}
\tau_{n}  = m^{2}_{f}e^{4A(\phi)} \tau_{n}\ . \label{detw}
\end{equation}
In region I, this reads
\begin{equation}
 -\frac{ d^{2} }{ d \phi^{2} } \tau_{n} = (
m_{n}^{2} e^{2H} - m_{f}^{2} ) \tau_{n}\ , \label{tequ}
\end{equation}
and it becomes in region II
\begin{equation}
 - \frac{ d^{2} }{ d \phi^{2} } \tau_{n} = (
m_{n}^{2}  - m_{f}^{2} ) \tau_{n}\ .
\end{equation}

  This system is in many ways similar to the square well
of one dimensional quantum mechanics. The solutions fall into two
categories, solutions analogous of bound states with
\begin{equation}
\left( m_{n}^{2} e^{2H} - m_{f}^{2} \right) > 0  \ \ \textnormal{but} \ \
 \left(m_{n}^{2} - m_{f}^{2}\right) < 0
\end{equation}
and the analogues of scattering states with
\begin{equation}
 \left(m_{n}^{2} - m_{f}^{2}\right) > 0\ .
\end{equation}

Let us first examine the bound states.
The equations of motion and the boundary
conditions require $\tau_{n} (\phi)$ to be of the form
\begin{equation}
\tau_{n} (\phi) = A \cos \left( \sqrt{
 e^{2H}m_{n}^{2} -m_{f}^{2} }
 \phi \right) \ \   \textnormal{(region I)}
\end{equation}
and
\begin{equation}
    \tau_{n}( \phi ) = B \cosh \left(\sqrt{
 m_{f}^{2}- m_{n}^{2}} (\phi - R )\right) \ \ \textnormal{(region II)}\ .
\end{equation}
Note that these modes have their wavefunction localized in region
I and exponentially suppressed in the bulk. Therefore, we call these
\textit{localized} modes.

          The masses $m_{n}$ are obtained from the boundary
conditions at the interface $\phi = L$.  Continuity of the
solution at the interface requires
\begin{equation}
 \tau_{n}(L) = A \cos \left( \sqrt{
 e^{2H}m_{n}^{2} -m_{f}^{2} }
 L \right) = B \cosh \left(\sqrt{
m_{f}^{2}- m_{n}^{2}} (L - R)\right)\ . \label{lc}
 \end{equation}
The matching condition for the derivative can be obtained by
integrating (\ref{de}) in a small neighborhood of the interface.
We see that $ e^{4A(\phi)}\tau'_{n}(\phi) $ must be continous;
that is,
\bea
&&  A e^{-4H} \sqrt{
e^{2H}m_{n}^{2} -m_{f}^{2}}\sin \left( \sqrt{
 e^{2H}m_{n}^{2} -m_{f}^{2} }
 L \right) = \nonumber\\
&&B \sqrt{m_f^2-m_n^2}\sinh \left(\sqrt{
m_{f}^{2}- m_{n}^{2}} (R - L)\right)\ . \label{dc}
\eea
The ratio
\bea
 &&e^{-4H} \sqrt{
 e^{2H}m_{n}^{2} -m_{f}^{2} }
   \tan \left( \sqrt{
 e^{2H}m_{n}^{2} -m_{f}^{2} }
 L \right)  =\nonumber\\
&&\sqrt{
 m_{f}^{2}- m_{n}^{2} } \tanh \left(\sqrt{
 m_{f}^{2}- m_{n}^{2}} ( R - L)\right) \label{nm}
\eea
of (\ref{dc}) and (\ref{lc})
then determines the masses. Although it is not possible to solve
(\ref{nm}) analytically, we can develop a fairly good understanding of the
spectrum by plotting the functions appearing on
both sides of the equation. We begin by defining the variable
\begin{equation}
 x = \sqrt{ e^{2H}m_{n}^{2} -m_{f}^{2} } \,L\ . \label{xde}
\end{equation}
The left-hand side of (\ref{nm}) then represents the curve
\begin{equation}
   y_{a}(x)  = e^{-4H} x \tan x\ ,  \label{tacr}
\end{equation}
and the right-hand side is
\begin{equation}
y_{b}(x) = \sqrt{ m^{2}_{f}L^{2}(1-e^{-2H})- e^{-2H}x^{2}
} \tanh \left[  \sqrt{
 m^{2}_{f}L^{2}(1-e^{-2H})- e^{-2H}x^{2}}  \left( \frac{R}{L} - 1\right)
\right]\ . \label{elcr}
 \end{equation}
\FIGURE[t]{
\includegraphics[scale=1]{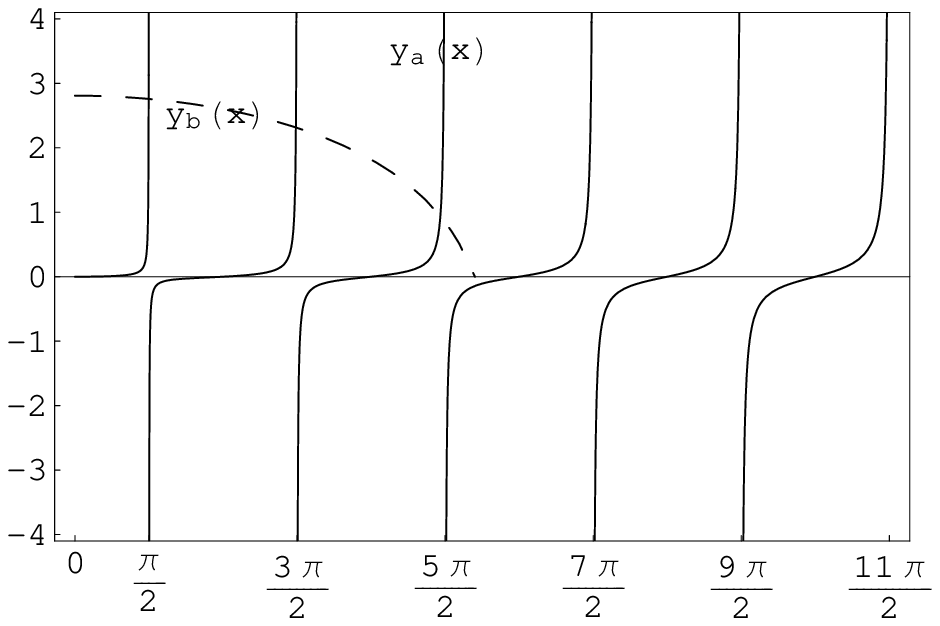}
\caption{\lb{f:bound} We can solve for the localized mode masses by finding
intersections of the solid ($y_a$) and dashed ($y_b$) curves.  The masses are
given in terms of the intersection points $x_n$ by equation
(\protect\ref{ma}). For illustrative purposes, we take $m_f L =3$,
$e^H=3$, and $R/L=2$.}}

  The curve $y_{a}(x)$ has a branch structure
characteristic of $\tan x$, so $y_a$ approaches infinity when $x$ approaches
odd multiples of $\pi/2$. On the other hand, $y_{b}(x)$ is
a squashed ellipse which intersects the $x$-axis at
\begin{equation}
 \hat{x} = L m_{f} \sqrt{e^{2H} - 1}\ . \lb{hx}
\end{equation}
   Hence there is one intersection in each interval of $\pi$
on the $x$-axis for $ x < \hat{x} $, and we have $ \lfloor
\hat{x}/\pi\rfloor  + 1$  localized modes.\footnote{Note
that, as in the square well problem in quantum mechanics, there is
always at least one localized mode.}

   Now we can analyze the spectrum using (\ref{xde}), which
can be written as
\begin{equation}
     m^{2}_{n} = e^{-2H} m^{2}_{f} \left[ 1 + \frac{ x_{n}^{2} } { m_{f}^{2}
     L^{2}  } \right]\ . \label{ma}
\end{equation}
For a deep throat ($e^{H} \gg 1$) with  $m_{f}^{2} L^2
\sim 1 $ (which is characteristic of throats in string
compactifications), the lowest mode has mass
\begin{equation}
    m_{0} \approx m_{f} e^{-H}\ . \label{mlow}
\end{equation}
The highest mass for a localized mode is given by $m_N\sim m_f$, simply
approximating $x_N\sim\hat x$
It is also important to consider the spacing of masses in order to determine
whether there is a hierarchy between the KK mass scale and the flux-induced
mass scale.  For large masses $m_n\sim m_f$, we can show that
\be\lb{deltamhi}
\Delta m_n \approx \frac{1}{L}\ ,\ee
which is the naive Kaluza-Klein scale of the throat.  However, for light KK
modes, near the lowest mass $m_0\sim e^{-H}m_f$, we find
\be\lb{deltamlow}
\Delta m_n \approx \frac{e^{-H}}{L}\ ,\ee
the \textit{warped} Kaluza-Klein scale.  In string compactifications, we
will see that $m_f$ and $L$ are set by approximately the same quantum
numbers and are larger than unity in string units,
so $\Delta m_n\lesssim m_n$ for small excitation number $n$.  This is a
crucial point for the validity of the flux superpotential in the 4D effective
field theory.  Also, we should note that the localized mode
masses are practically independent of $R$, the volume
of the bulk region, depending on it only weakly through $x_n$.

      Next, we analyze the analogues of scattering states.
For these, the Kaluza-Klein wavefunction takes the form
\begin{equation}
 \tau_{n} (\phi) = A \cos \left( \sqrt{
 e^{2H}m_{n}^{2} -m_{f}^{2} }
 \phi \right)
\end{equation}
in region I and
\begin{equation}
 \tau_{n}(\phi) = D\cos \left( \sqrt{
 m_{n}^{2} -m_{f}^{2} }
 (\phi - R ) \right)
\end{equation}
in region II.
These solutions have an oscillatory profile in the bulk, so we
call them \textit{bulk} modes.

Again,  the matching
conditions at $\phi = L$ determine the masses:
 \begin{equation}
 e^{-4H} x \tan x = \sqrt{ e^{-2H}x^{2} -
 m^{2}_{f}L^{2}(1-e^{-2H})} \tan \left(  \sqrt{ e^{-2H}x^{2} -
 m^{2}_{f}L^{2}(1-e^{-2H})}  \left( 1 -\frac{R}{L}\right)
\right) \label{crct}
\end{equation}
($x$ is defined in terms of the mass as before).  The
left-hand-side of (\ref{crct}) is just the curve
$y_{a}(x)$ defined in (\ref{tacr}). We add $y_{c}(x)$, a curve
representing the right-hand-side of (\ref{crct}) to figure \ref{f:bound}.
Note that the condition $m^{2}_{n} > m^{2}_{f} $ requires us
to consider only $ x > \hat{x} $.  See figure \ref{f:allmodes}.

\FIGURE[t]{
\includegraphics[scale=1]{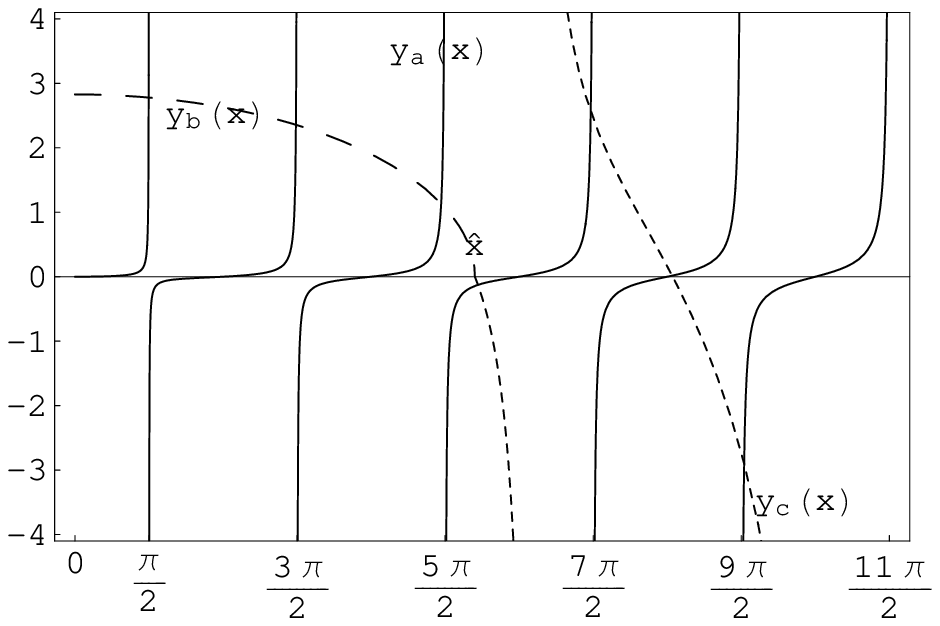}
\caption{\lb{f:allmodes} Plot showing all states in our example.
The intersections of the dotted curve ($y_c$) with the solid curve ($y_a$) give
the masses of the scattering modes.  $\hat x$ indicates
the cross-over between localized and bulk modes.
The parameters of the model are as
in figure \protect\ref{f:bound}.}
}

        The spectrum of both localized and bulk modes can be
inferred from figure \ref{f:allmodes}. We would like to point out
one important feature. Consider increasing the warped hierarchy
$H$ continuously, so the point $\hat{x}$ toward the right along
the $x$-axis. As a result, more and more bulk modes are smoothly
converted to localized modes. The wavefunctions of the lower mass
modes become highly localized in the region of significant
warping. Also note that, for $m_{f} = 0$, there is a mode of zero
mass and constant profile for all values of $H$. As $m_{f}$
increases from zero, the wavefunction of this mode gradually
localizes (for non-zero $H$). The wavefunction of this lowest mass
mode can always be written as a linear superposition of the
eigenmodes at $m_{f}= H = 0 $ (since these form a complete
set).\footnote{These are just the naive kk modes and have the same
wavefunctions for nonzero $m_f$ and $H=0$.} Since the lowest mode
at $m_{f},H=0$ has a constant profile, any deviation of
$\tau_0(\phi)$ from a constant should be regraded as a mixing of
the naive kk modes. Note that this mixing is energetically
\textit{favorable}, since it localizes the wavefunction in a
region of large warping.

The idea of kk mode mixing suggests a treatment in perturbation
theory. We can imagine two types of perturbation: perturbing
around an unwarped background with finite $m_f$ (that is, $Le^H\to
0$) or adding a perturbative mass $m_f\to 0$ to a warped
background ($Le^H$ finite).  As it turns out, both of these
perturbations are mathematically equivalent. For $m_{f}^{2} L^{2}
e^{2H}\ll 1$ (and $R/L\to\infty$ for simplicity),
\begin{equation}
   x^{2}_{0} \approx m_{f}^{2} L^{2} \left(e^{2H}-1\right)
\left[1-\frac{1}{2} e^{-6H} m_f^2 L^2  \left(e^{2H}-1\right)
\right]
\end{equation}
and
\begin{equation}
    m_{0} \approx m_{f} \left[ 1-\frac{1}{4} m_f^2 L^2
e^{-8H} \left(e^{2H}-1\right)^2\right]\ .\label{ptb}
\end{equation}
Even perturbatively, we see that the lightest mode becomes localized
and has its mass reduced.
\comment{At the moment, it is clear how to derive (\ref{ptb}) from perturbation
theory around $m_f=0$, since we can map the equation of motion to a
Schr\"odinger equation (see section \ref{s:simpleeom} for a similar
example).  Unfortunately, due to the unusual boundary conditions at
$\phi=L$, it is unclear how to phrase the perturbation theory around
$L=0$.}

Finally, the astute reader may have noticed that we did not include the
possibility of modes with $m_n e^{H}<m_f$.  These would be supported by
the matching conditions at $\phi=L$ in much the same way that a
delta-function potential can support a bound state and the same as the
origin of the graviton zero mode in an RS compactification.  However,
the matching condition in that case would be
\be\lb{match-delta}
-e^{-4H} |x|\tanh|x| = \sqrt{(1-e^{-2H})m_f^2+e^{-2H}|x|^2}
\tanh\left[ \sqrt{(1-e^{-2H})m_f^2+e^{-2H}|x|^2} \left(\frac{R}{L}-1
\right)\right]\ ,\ee
where $x$ as defined in (\ref{xde}) is now imaginary.  Equation
(\ref{match-delta}) has no solutions because the two sides have opposite
sign, so there are no very light bound states.  Note the contrast with
the case of the graviton.

\section{Geometry and Flux}\label{s:geometry}

          In this section we review the compactifications
of type IIB string theory constructed in \cite{hep-th/0105097}.
Our treatment will be very brief, focusing on some specific
features useful for our later discussion. For detailed and
comprehensive reviews see \cite{hep-th/0308156,hep-th/0405068,hep-th/0509003}.

In these constructions, the geometry\footnote{We shall 
use the 10D Einstein frame for
our discussion and follow the conventions of \cite{hep-th/0105097}.} 
takes the form of a warped product
\begin{equation}
ds_{10}^{2} = [e^{-4A} + c]^{-1/2} \eta_{\mu \nu} dx^{\mu}
dx^{\nu} + [e^{-4A} + c]^{1/2} \tg_{mn}(y) dy^{m} dy^{n}
\label{gkpm}
\end{equation}
where $\tg_{mn}$ is the metric of a unit Calabi-Yau.  The overall
size of compactification is controlled by the parameter $c$ (for
details see \cite{hep-th/0507158}). For convenience, we will also
denote $e^{-4\t A}=c+e^{-4A}$ throughout this paper.  In addition, there
are nonzero components of R-R ($F_3$) and NS-NS ($H_3$) flux wrapping
3-cycles of the Calabi-Yau manifold.  In the background compactification,
$G_3=F_3-\tau H_3$ must be imaginary-self-dual on the Calabi-Yau, and
this condition generically freezes the complex structure moduli as well
as the dilaton-axion $\tau=C+ie^{-\phi}$.

                                    The warp factor
$e^{-4A}$ satisfies a Poisson equation sourced by D3-brane
charge and 3-form flux
\begin{equation}
    - \tilde{\nabla}^{2} e^{-4A} =  \frac{G_{mnp}
    \bar{G}^{\widetilde{mnp}}}{12 \im \tau } + 2 \kappa_{10}^{2} T_{3}
\tilde{\rho} _{3}^{\mathnormal{loc}}\ ,
    \label{wfac}
\end{equation}
where the tilde indicates use of the underlying Calabi-Yau metric.
   In most parts of the manifold, the warpfactor is approximately
constant and the background resembles that of a standard
Calabi-Yau compactification. This Calabi-Yau-like region is often
referred to as the \textit{bulk}. We have chosen conventions for
the warp factor such that $e^{-4A}\to 0$ in the bulk; this
convention is useful in that it makes clear the dependence of
warping-induced hierarchies on the bulk volume modulus, of course
the physically meaningful warp factor is given by $\t A$. We also
note that (\ref{wfac}) requires some source of negative D3-brane
charge, which can be provided by O3-planes, O7-planes, or F theory
compactifications.

 The reader should note that the metric (\ref{gkpm}) is not in a
standard frame, since the external Minkowski metric is multiplied
by a nontrivial constant in the bulk Calabi-Yau region. The
massless graviton gives rise to a metric $\hat g_{\mu\nu}$ defined
via \be\lb{rescaling} [1+c^{-1} e^{-4A}]^{-1/2}\hat g_{\mu\nu} =
e^{2\t A}g_{\mu\nu}\ \Rightarrow\ \ \hat g_{\mu\nu} \approx
c^{-1/2}g_{\mu\nu}\ .\ee We are used to dealing with the masses
measured by $\hat g_{\mu\nu}$, so the masses in this paper will be
rescaled compared to our normal intuition by $m=c^{-1/4}\hat m$.
In other words, while the conventional mass of a Kaluza-Klein mode
with trivial warping $( e^{-4A} = 0 )$ is $\hat m \sim
1/c^{1/4}\sqrt{\ap}$, we will have a Kaluza-Klein mass scale of $m
\sim 1/\sqrt{c\ap}$. See also the discussion in \cite{hep-th/0507158}.

In the bulk region, the flux is approximately constant at large
volume. We define
\begin{equation}\lb{bulkg}
  \frac{1}{12} G_{mnp}\bar{G}^{mnp} = \frac{1}{12} e^{6\t A}G_{mnp}
\bar{G}^{\widetilde{mnp}}
\approx \frac{n'^{2}_f}{c^{3/2} \alpha'}\ .
\end{equation}
The explicit dependent on $c$ arises
due to the scaling of the metric with the overall volume.
We have defined $n'_f$ as a sum over cycles of squares of flux quantum
numbers,
\be\lb{n'}
n'^{2}_f = \sum \left| n_{RR}-\tau n_{NS}\right|^2\ .\ee
In a general Calabi-Yau, $|G|^2$ will have some position dependence,
but we expect the flux to be approximately
constant except near shrinking 3-cycles.  However, shrinking 3-cycles
with wrapped flux will give rise to the throat regions described below,
which we will described separately from the bulk.
The norm of the flux is important to us, as it
induces a mass for the frozen moduli, including $\tau$.

In addition to the bulk of the Calabi-Yau, small regions of large
($e^{-4A}\gg c$)
warping can arise from typical values of the flux quantum
numbers, as shown in \cite{hep-th/0105097}.
For instance, if  $M$ units of R-R flux
and $K$ units of NS-NS flux wrap the A and B cycles of a conifold
locus respectively, the warp factor is exponential in the ratio of
flux quanta:
\begin{equation}\lb{wconifold}
       e^{-4A_{0} } \approx e^{8 \pi K/3M g_{s} }\ .
\end{equation}
Such warped regions are called \textit{throats}. The four
dimensional energy of phenomenon localized in these regions is approximately
redshifted by a factor of $e^{A_{0}}c^{1/4} $ compared to modes in
the bulk, making throats attractive for a whole host of
phenomenological applications as indicated in
\cite{hep-ph/9905221,hep-th/0208123}. 
This redshift factor is actually the ratio of $e^{\t A}$
in the throat and the bulk; note that the redshift becomes of
order one for $c \sim e^{-4A_0}$. For larger $c$, the throat
disappears entirely because $e^{-4A}$ contributes only a small
perturbation around the value $e^{-4\t A}\sim c$.  Therefore, we refer
to the regime $c\gg e^{-4A_0}$ as the infinite volume limit.

       The local geometry in throats is of much interest
to us. In these regions, the metric becomes independent of $c$, and
the geometry is similar to that of the solution found in \cite{hep-th/0007191}
(henceforth KS). The KS solution is a solution of the form
(\ref{gkpm}) in which
the Calabi-Yau $\tg_{mn}$ in (\ref{gkpm}) is the non-compact
deformed conifold.  We note some features
of this solution essential for our analysis and refer the reader to
\cite{hep-th/0007191,hep-th/0108101,hep-th/0205100,hep-th/0002159}
for details. In this case, it is possible to solve for the warp
factor explicitly. The warp factor $e^A$ depends only on
the radial direction of the conifold, and it attains its minimum
value on the finite size $S^{3}$ associated with the conifold deformation.
The minimum value, which is that given in (\ref{wconifold}),
is such that the proper size of the $S^{3}$ in the warped metric is
independent of the deformation parameter of the conifold.  The radius of
the $S^3$ is approximately $R_{S^3}\sim g_s^{1/4}\sqrt{M\ap}$.

In regions away from bottom of the throat, the geometry takes the
form of a product $\mathnormal{AdS}_{5}\times T^{1,1}$ (up to
logarithmic corrections in the warp factor), so
\begin{equation}
    ds^{2} = \frac{r^{2}}{R^{2} } \eta_{\mu \nu}dx^{\mu}
    dx^{\nu} + \frac{R^{2}}{r^{2}} dr^{2} +
    R^{2} ds^{2}_{T^{1,1}}\ . \label{appr}
\end{equation}
The $\mathnormal{AdS}$ radius $R$ (near which the throat connects
to the bulk) is given by the effective three brane charge, $R^{4}
\sim MK \ap{}^2$. We model the KS throat by the metric
(\ref{appr}) by restricting ourselves to $r> r_{0}$ with \be
    \frac{r_{0}}{R} = e^{A_{0}}
\ee
and choosing appropriate boundary conditions at $ r =r_{0}$ for
fluctuating modes. This geometry with flux (and logarithmic corrections)
was first discussed by Klebanov and Tseytlin \cite{hep-th/0002159} 
(henceforth KT).
We will restrict our attention to this simplified KT model.

Let us now consider the flux in a KS throat.  For a noncompact
deformed conifold, the only flux components wrap the conifold A
and B cycles. Just like in the bulk, the quantity
$G_{mnp}\bar{G}^{mnp}$ is approximately constant in a KS throat
(upto logarithmic corrections) for the A and B cycle fluxes, which
follows directly from the solution for the flux in a KS
background. This flux is also independent of the volume of the
compactification due to the fact that geometry of the throat is
practically independent of the bulk volume.  However, with a
compact Calabi-Yau attached to the KS throat, there are other
possible flux components.  These may have nonconstant $|G|^2$ in
the throat, but we can assume that their norm is either constant
or decreases in regions of small warp factor.  Otherwise, they
would alter the KS geometry significantly, possibly leading to a
naked singularity. Therefore, as in the bulk case (\ref{bulkg})
above, we will define
\be\lb{ksg} \frac{1}{12}G_{mnp}\b G^{mnp}=\frac{1}{12} 
e^{6\t A}G_{mnp}\b G^{\widetilde{mnp}} =
\frac{n_f^2}{\ap}\ , \ee
where $n_f^2$ is a constant measuring the
number of flux quanta (defined as in (\ref{n'}) above).

There are also other types of throats corresponding to flux
wrapping cycles near other types of singularities, as described in
\cite{hep-th/0502113}. In these throats, the metric takes the
form (\ref{appr}) with $T^{1,1}$ replaced by another
Einstein-Sasaki manifold far from the bottom of the throat.  The
bottom of these throats (that is, the region with smallest warp
factor) are expected to have a geometry similar to the bottom of
the KS throat.  Further, from studying the solutions in these
throats, we see that we can write the flux norm as (\ref{ksg}).
Therefore, our results will apply to general throat regions in
this class of compactification.

We close our review with a word on moduli stabilization.  As we
mentioned above, the requirement that $G_3$ be imaginary-self-dual
on the internal space generically implies that $\tau$ and the
Calabi-Yau complex structure moduli must be fixed to specific
values.  Correspondingly, fluctuations around those vacuum
expectation values are massive; in the 4D effective theory, it can
be argued that those masses arise due to a superpotential \be
\lb{gvw} W \propto \int G_3\wedge\Omega\ , \ee where $\Omega$ is
the holomorphic $(3,0)$ form on the Calabi-Yau.  From this
superpotential, we see immediately that the frozen moduli must
have masses $m^2\sim |G|^2$. In the following section, we will see
from the linearized 10D equations of motion that a natural choice
of mass scale is what we will call the ``flux-induced mass''
\be\lb{mfdef} m_f^2(y) \equiv \frac{g_s}{12} G_{mnp}\bar G^{mnp}\
,\ee 
which generally will have some dependence on position in the
compactification. In particular, $m_f^2\propto n_f^2$, where
$n_f^2$ is the summed square of flux quantum numbers (as in
(\ref{n'})), which may differ in bulk and throat
regions.\footnote{In our frame, the mass of a frozen modulus in
the infinite volume limit is actually $m_f/c^{1/4}$ ; $m_f$ is
instead the mass in the conventional frame.}  In a throat region,
we have $m_f\sim 1/\sqrt{M\ap}$, which is about unity in
$\mathnormal{AdS}$ units, $m_f R\sim 1$, and is slightly suppressed
compared to the string scale.  Also, for
typically Calabi-Yau manifolds, $M$ cannot be too large due to a
Gauss constraint on D3-brane charge. Finally, we note that
\cite{hep-th/0105097} showed that the dilaton-axion and conifold
deformation modulus cannot both be stabilized by the KS throat
flux --- additional components are necessary.  In section
\ref{s:detailed}, we will consider two different models for
$m_f(y)$ to take this fact into account.

\section{Linearized Equations of Motion}\label{s:eom}

In this section, we will first study the full linearized equations of
motion for fluctuations of the dilaton-axion, arguing that they can
be simplified to a lowest approximation.  Then we give the simplified
differential equation that we will take to govern the dilaton-axion's
Kaluza-Klein wavefunction.

\subsection{Compensator Equations}\label{s:compensators}
As was emphasized in \cite{hep-th/0507158} (and to a lesser
extent in \cite{hep-th/0201029,hep-th/0308156}), the equations
of motion of type IIB SUGRA mix the perturbative modes of
many supergravity fields.  Simply put, the field that corresponds
to the dilaton-axion in the 4D EFT contains excitations of
the 10D dilaton-axion along with many other SUGRA modes.
The reason is that the dilaton-axion fluctuation appears
as a source in the linearized equations of motion for the other SUGRA
fields, which leads to constraint equations beyond the dynamical
equation of motion.  We are therefore forced to allow fluctuations
of other SUGRA fields, which act as ``compensators'' for the dilaton-axion
fluctuation.  In this section, we will study the compensator
equations for the dilaton-axion and argue that they give small corrections
to the dilaton-axion dynamics.

The type IIB SUGRA equations of motion in 10D Einstein frame are
\bea
\Del^2 C &=& -2\del_M\phi\del^M C-\frac{1}{6} e^{-\phi} \t F_{MNP}H^{MNP}
\lb{dilax01}\\
\Del^2 \phi &=& e^{2\phi}\del_M C\del^M C
+\frac{1}{12} e^{\phi} \t F_{MNP}\t F^{MNP}-\frac{1}{12} e^{-\phi}
H_{MNP}H^{MNP}\lb{dilax02}\\
d\t F_5&=& H_3\wedge F_3=H_3\wedge \t F_3\lb{5form}\\
d\star\left(e^\phi \t F_3\right)&=& \t F_5\wedge H_3 \lb{r3}\\
d\star\left(e^{-\phi}H_3-e^\phi C\t F_3\right)&=&-\t F_5\wedge F_3
\lb{ns3}\eea
along with the Einstein equations.  We will proceed by linearizing these
equations, starting with the dynamical equations (\ref{dilax01},\ref{dilax02})
for the dilaton-axion.  We will demonstrate that
the 10D dilaton-axion cannot be the only fluctuations in a background with
imaginary self-dual $G_3=F_3-\tau H_3$.
Let us consider an orientifold background for our flux compactification,
in which the background dilaton-axion $\tau=C+i/g_s$ is constant.  With
O3 and O7 orientifold planes, the dilaton-axion, metric, and
4-form potential fluctuations have
even boundary conditions at the O-planes, while both of the 2-form potentials
have odd boundary conditions.

We define
\be\lb{deltaG}
\delta\hat F _3 = \delta F_3-C \delta H_3\ ,\ee
and we also note that
\be\lb{selfdual}
\t F_{mnp} H^{mnp} = 0\ \textnormal{and}\ \frac{1}{2}G_{mnp}\bar G^{mnp}=
\t F_{mnp}\t F^{mnp}= \frac{1}{g_s^2} H_{mnp}H^{mnp}\ee
due to the imaginary self-duality of the complex three form on internal
space.  In this type of background, the linearized equations of motion for
the dilaton-axion are
\bea
\Del^2 \delta C &=& \frac{g_s}{12}\delta C G_{mnp}\b G^{mnp} -\frac{1}{6g_s}
\left(\delta \hat F_{mnp}H^{mnp} +\delta H_{mnp}\t F^{mnp}\right)\nonumber\\
&& -\frac{1}{2g_s} \delta g_{mn}\t F^m{}_{pq}H^{npq}\nonumber\\
\Del^2\delta\phi &=& \frac{g_s}{12}\delta \phi G_{mnp}\b G^{mnp}
+\frac{g_s}{6}
\left(\delta \hat F_{mnp}\t F^{mnp} +\frac{1}{g_s^2}\delta H_{mnp}H^{mnp}
\right)\nonumber\\
&&+\frac{g_s}{4}\delta g_{mn}\left(\t F^m{}_{pq} \t F^{npq}-\frac{1}{g_s^2}
H^m{}_{pq}H^{npq}\right)\ .\lb{dilax1}\eea
Here $\delta g_{mn}$ is the complete
change in the 6D metric, including any fluctuation in the warp factor, and
the second line of each equation above indicates tree-level mixing between
the dilaton-axion and internal metric fluctuations.  This mixing
appears in models with small numbers of flux
components, for example \cite{hep-th/0201028,hep-th/0210254}.  
From this point, we will neglect such
mixing for simplicity.

Finally, splitting the Laplacian between internal and external dimensions
and accounting for the warp factor,
\be\lb{dilax3}
\left(\t\Del^2 +e^{-4\t A}m^2-\frac{g_s}{12}e^{4\t A} G_{mnp}
\b G^{\widetilde{mnp}}\right)
\left[\begin{array}{c}\delta C\\ \delta\phi\end{array}\right]
= e^{4\t A}\left[\begin{array}{c}
-\frac{1}{6g_s}\left(\delta\hat F_{mnp}H^{\widetilde{mnp}}
+\delta H_{mnp}\t F^{\widetilde{mnp}}\right)\\
\frac{g_s}{6}\delta\hat F_{mnp}\t F^{\widetilde{mnp}}
-\frac{1}{6g_s}\delta H_{mnp}H^{\widetilde{mnp}}\end{array}\right]\ ,
\ee
where $m$ is the mass in the 4D EFT.
Tildes indicate use of the underlying Calabi-Yau metric $\t g_{mn}$.
The left-hand side of this equation (when set to zero)
is the naive equation of motion for
the dilaton-axion zero-mode in 3-form flux.
The right-hand side
indicates mixing with fluctuations of the 3-form flux, even though those
fluxes are not naive KK zero-modes due to the orientifold projection.

With a trivial warp factor, the dilaton-axion zero-mode is the
naive kk zero mode (that is, constant), and it does not mix with
the 3-form fluctuations.\footnote{We are approximating the norm of
the flux $G_{mnp}\t G^{mnp}$ to be constant on a large volume
Calabi-Yau.} However, with nontrivial warping, the flux term on
the left-hand side is position dependent, even on a torus with
constant flux.\footnote{However, it is possible to use the supersymmetry
variations to argue
that the dilaton-axion zero mode is constant \textit{for any warp
factor} in the absence of flux.} Therefore, the naive kk modes of
the dilaton-axion must mix with each other, and the dilaton-axion
further mixes with naive kk excitations of the 3-form flux. In
that light, we should examine the other equations of motion. The
5-form equation linearizes to \be\label{5form1} d\delta\t F_5 =
\delta H_3\wedge \t F_3+ H_3\wedge\delta\t F_3\ ,\ee so the 5-form
perturbations are sourced if the 3-form perturbations are. Without
writing out the full Einstein equations, we can also say that the
warp factor must also be modified at the linear level in order for
dilaton-axion perturbations to solve the equations of motion.

We will look at the 3-form equations in a little more detail.  Using the
fact that the background $\tau$ is constant, we can simplify the linearized
equations to
\bea
d\star\left(-g_s^{-1} \delta\phi H_3 +g_s^{-1} \delta H_3-g_s\delta C
\t F_3\right)+g_s^{-1}d(\delta\star)H_3&=& -\delta\t F_5\wedge\t F_3
-\t F_5 \wedge\delta\hat F_3 \nonumber\\
g_s d\star\left(\delta F_3-C\delta H_3-\delta C H_3\right)+g_s
d(\delta\star)\t F_3
&=&
\delta\t F_5\wedge H_3+\t F_5\wedge\delta H_3\ .\lb{3form1}
\eea
The change in the Hodge duality rule ($\delta\star$) embodies both
fluctuations in the unwarped metric and the warp factor.  Every possible
fluctuation is involved in these equations, and there is not a simple
solution, just as has been seen in other cases
\cite{hep-th/0201029,hep-th/0308156,hep-th/0507158}.

In order to estimate
the effect of the compensators on the dilaton-axion zero-mode, we
consider a simplified system.  We stress that we are not attempting to
solve the compensator equations but rather are studying a truncation of
them as a rough guide to their effects.
First, we take all fluctuations to vanish
except for $\delta\tau$, $\delta F_3$, and $\delta H_3$. This choice is
appropriate because the 3-form fluctuations appear directly in the
dilaton-axion equations (\ref{dilax3}).
With some massaging, we can then write (\ref{3form1}) as
\bea
\star d\star\delta \hat F_3  -(d\delta C)\lrcorner
H_3 &=& -\frac{1}{g_s}\delta H_3 \lrcorner \t F_5\nonumber\\
\star d\star\delta H_3-2(d\delta\phi)\lrcorner H_3 -
g_s^2(d\delta C)\lrcorner \t F_3 &=& g_s\delta \hat F_3\lrcorner \t F_5
\ ,\lb{3form2}
\eea
where
\be\lb{contract}
\left(\alpha_p\lrcorner \beta_q\right)_{N_1\cdots N_{q-p}}
=\frac{1}{p!}\alpha^{M_1\cdots M_p}\beta_{M_1\cdots M_P N_1\cdots
N_{q-p}}\ .\ee

The next simplification we consider is to specify the geometry we consider;
we will take a KS throat approximated by the KT metric (\ref{appr}) of
$\mathnormal{AdS}_5\times T^{1,1}$ and further approximate $T^{1,1}$ as
$S^3\times S^2$.  We can reduce the dilaton-axion equation (\ref{dilax3})
to a 5D
by taking the dilaton-axion to depend only on the external directions
and the radial dimension $r$.  Since we want to keep the system of equations
5D, we should also take the 3-form fluctuations constant on the spheres.
Then spherical symmetry suggests that we take
\be\lb{2forms}
\delta B_{\theta\phi}\left(x^\mu, r\right)\ ,\ \
\delta C_{\theta\phi}\left(x^\mu, r\right)\ee
as the only nontrivial components of the NS-NS and R-R 2-form potentials.
Here $\theta,\phi$ are the coordinates of the $S^2$.
Finally, we truncate the equations of motion to (\ref{dilax3}) and
the $\theta,\phi$ component of (\ref{3form2}).
In this simplified background, there are a number of cancellations, and
the equations of motion become (in the NS-NS sector)
\bea
\left(\del_r^2 +\frac{5}{r}\del_r-m_f^2\frac{R^2}{r^2}
+m_\phi^2 \frac{R^4}{r^4}\right)\delta\phi&=& -\frac{1}{6g_s}
\frac{n_h}{r}\del_r
\delta\hat B\nonumber\\
\left(\del_r^2 +\frac{5}{r}\del_r+m_B^2\frac{R^4}{r^4}\right)\delta \hat B
&=& \frac{n_h}{R}\frac{r}{R}\del_r\delta\phi\ ,
\lb{coupledns}
\eea
and the R-R fields satisfy the similar equations
\bea
\left(\del_r^2 +\frac{5}{r}\del_r-m_f^2\frac{R^2}{r^2}
+m_\phi^2 \frac{R^4}{r^4}\right)\delta C&=& -\frac{1}{6g_s}\frac{n_h}{r}\del_r
\delta\hat C\nonumber\\
\left(\del_r^2 +\frac{5}{r}\del_r+m_B^2\frac{R^4}{r^4}\right)\delta \hat C
&=& \frac{n_h}{R}\frac{r}{R}\del_r\delta C\ .
\lb{coupledrr}
\eea
Here, the number $n_h$ is defined by $H_{r\theta\phi}=n_h R^2/r$, which is the
appropriate behavior for the NS-NS flux in a KS throat, and we have
written the shorthand $m_f^2$ as in (\ref{mfdef}).
We have also defined
$\delta\hat B = \delta B_{\theta\phi}/R^2$ and
$\delta\hat C = \delta C_{\theta\phi}/R^2$ to make all our fields
dimensionless.

We have solved the equations (\ref{coupledns}), treating the right-hand-sides
as perturbations.  Our results show that this approach is self-consistent;
for example, when the full hierarchy of the KS throat is $10^{-3}$ and
reasonable values are taken for compactification parameters, the change
in mass-squared of the dilaton is about 1 part in $10^4$.  Furthermore,
from our numerical estimates, it appears that $|\Delta (m_\phi^2)|
\sim e^{A_0} m_\phi^2$, where $e^{A_0}$ is the minimum warp factor
of the throat.  Note that we have assumed that $c=0$ in modeling the
throat as $\mathnormal{AdS}_5\times T^{1,1}$; in a more realistic
compactification, we expect the compensators to give a negligible contribution
in the bulk region.  See the discussion above equation (\ref{5form1}).
See appendix \ref{s:compapp} for more details of our estimate.

\subsection{Simplified Equation of Motion}\lb{s:simpleeom}

For simplicity, we approximate that the internal metric factorizes into a
radial direction and 5D part, much as in the
$\mathnormal{AdS}_5\times T^{1,1}$ KT geometry, and we take
\be\label{6dmetric}
d\t s^2 = dr^2 +L^2[e^{-4A}+c]^{-1/2} ds_5^2
=dr^2+L^2 e^{2\t A}ds_5^2\ ,\ee
where $ds_5^2$ may be defined piecewise along the $r$ direction.  Here,
$L$ is the linear scale of the 5 ``angular'' dimensions; in the
$\mathnormal{AdS}_5\times T^{1,1}$ background, $L=R$, the $\mathnormal{AdS}$
radius.

The dilaton-axion equation of motion involves the scalar Laplacian
of $\t g_{mn}$, which is
\be\lb{laplace}
\t\Del^2 \Psi = \left( \del_r^2 +5\del_r \t A \del_r +\frac{1}{L^2}e^{-2\t A}
\Del_5^2\right)\Psi\ee
on any scalar $\Psi$.  We can find this relation simply using formulae for
conformal scaling of the metric.  Therefore, ignoring compensators,
which give only a small correction, the dilaton-axion equation of motion
becomes\footnote{This decomposition figures into the calculation of
(\ref{coupledns},\ref{coupledrr}) as well.}
\be\lb{dilax-r}
\left(\del_r^2 +5\del_r \t A \del_r -e^{-2\t A}\frac{Q^2}{L^2} -e^{-2\t A}
m_f^2+e^{-4\t A}m^2\right)\delta\tau=0\ .\ee
Here, $Q^2$ is defined as eigenvalue of the fluctuation with respect to the
5D Laplacian.  Since we have oversimplified the full compactification
geometry, it may be necessary to define $A$ and $Q$ in a piecewise manner
over the range of $r$.  Also, $m_f$ is a function of $r$ in the most
general situation.  However, if the warping is just a Randall-Sundrum
($\mathnormal{AdS}_5$) throat, the differential equation
reduces to the left-hand-side of (\ref{coupledns}) above, with the
addition of angular momentum:
\be\lb{basic}
\left[\del_r^2 +\frac{5}{r}\del_r-\left(m_f^2+\frac{Q^2}{L^2}
\right)\frac{R^2}{r^2}+m^2 \frac{R^4}{r^4}\right]\delta\tau=0
\ .\ee
At points where the definition of the warp factor or angular momentum
are discontinuous, (\ref{dilax-r}) yields the boundary conditions
\be\lb{bdry-r}
\delta\tau\ ,\ \ e^{5\t A}\del_r\delta \tau \ \ \textnormal{continuous}\ .\ee

It is useful to recast the equation of motion (\ref{dilax-r}) into
a Schr\"odinger-like form, as has been done in the Randall-Sundrum
case for graviton KK modes 
\cite{hep-ph/9905221,hep-th/0001033,hep-th/0006191}. It is
a simple matter to write \be \lb{dilax-x} \left( -\del_x^2
+V(x)\right)\psi(x) = m^2 \psi(x)\ee for \be\lb{redefs-x} dx = \pm
e^{-2\t A}dr\ ,\ \ \delta\tau=e^{-3\t A/2}\psi\ , \ee and
potential \be\lb{dilax-V} V(x) = e^{-3\t A/2}\del_x^2e^{3\t
A/2}+e^{2\t A} \left(m_f^2 +\frac{Q^2}{L^2}\right)\ . \ee Note
that the sign and integration constant in the definition of $x$
(\ref{redefs-x}) may be chosen for convenience (and to make $x$
continuous if the warp factor is defined only piecewise
continously).

           To build some intuition,  we begin by considering a
simplified KT model for the fluctuations of the dilaton. We consider its
propagation in a finite domain $r_0<r<r_c$
of $ AdS_{5} \times T^{1,1} $,
with $r_{0}/R = e^{A_{0}}$ and $r_{c}^{2}/R^{2}= 1/\sqrt{c}$.
Then we have $x=R^2/r$ and a hierarchy in the throat of
$e^H\equiv x_0/x_c = e^{-A_0}/c^{1/4}$.
We incorporate the effects of the infrared throat region
and the bulk by imposing Neumann boundary conditions on $\delta\tau$
at both ends $r_0,r_c$. In this approximation, we effectively have a RS
model with five additional dimensions. For simplicity, we
take $m_{f}$ to be a constant. Our treatment, therefore, just
becomes mathematically equivalent to that of Goldberger and Wise 
\cite{hep-ph/9907218}.
(We shall address the issue of
varying flux in sections \ref{s:semiclassical} and \ref{s:detailed}.)
The potential is
\begin{equation}
   V(x) = \frac{1}{x^2}\left(\frac{15}{4} + m_f^2R^2
+Q^{2} \right)\ .
   \label{adsp}
\end{equation}
With this potential, the Schr\"odinger equation (\ref{dilax-x}) is
a well-known form of the Bessel equation, with (unnormalized) solution
\be\lb{bessel-ads} \psi(x) = \sqrt{\frac{x}{R}}J_{\nu}(mx)
 + b_{\nu} \sqrt{\frac{x}{R}}Y_{\nu}(mx)\ ,\ \ \nu = \sqrt{4+ Q^2
+m_f^2R^2}\ .\ee
(Note that this solution is only valid for a nonzero mass, $m\neq 0$.)
The boundary conditions can then be used to obtain an equation for
the masses,
\begin{eqnarray}
  \left[2J_{\nu}(z_{n})+ z_{n } J'_{\nu}(z_{n}) \right]\left[
2 Y_{\nu}\left( z_{n} e^{-H}\right) + z_{n}
e^{-H} Y'_{\nu}\left( z_{n}e^{-H}\right)\right] -&&
   \nonumber \\
\left[ 2 Y_{\nu}(z_{n}) + z_{n} Y'_{\nu}(z_{n})\right]\left[2 J_{\nu}\left(
z_{n} e^{-H}\right) + z_{n}
e^{-H} J'_{\nu}\left( z_{n}e^{-H}\right)
\right] &=& 0\ ,\lb{adsbc}
\end{eqnarray}
where  $ z_{n} = x_0 m_{n}$.  For large hierarchy ($H\gg 1$), the
condition (\ref{adsbc}) simplifies to
\be\lb{spectral}
z_{n}^{-\nu} \left(2J_{\nu}(z_{n})+ z_{n} J'_{\nu}(z_{n})
\right)=0 \ee
 for the lowest mass modes. Note that the prefactor,
which arises from the Bessel $Y_\nu$ functions, prevents $z=0$
from being a root, which is consistent because the Bessel function
solution is not valid for a massless mode.  (When $Q=0$ and
$m_f=0$, we know there should be a massless mode; this mode arises
from the solution $\psi\propto x^{-3/2}$ --- $\delta\tau$ constant
--- for the potential (\ref{adsp}).  As shown in \cite{hep-ph/9907218}, this
lowest mode gains a mass in perturbation theory as $m_f$ increases
from zero.)

The lowest mass mode has $Q=0$. As it turns out, the lowest root of
(\ref{spectral}) is of the order of $(\nu - 2)$, so
\begin{equation}
      m_{0} \approx e^{A_0} m_f=\frac{r_{0}}{R} m_{f}\ . \label{adsm}
\end{equation}
The reader may question how we know that the mass of this lowest mode
is proportional to
$m_f$.  Actually, this is straightforward to check numerically for
moderate values of the flux and large hierarcy
(for example, we can calculate
the derivative of the mass with respect to $\nu$).  Also,
the profile of the lowest-mass mode is highly localized in the region
close to $r_{0}$.

The separation of the roots of (\ref{spectral}) is also of the order
of unity; hence, the mass scale of the Kaluza-Klein tower is
\begin{equation}
   \Delta m_{KK} \approx  \frac{r_{0}}{R^{2}}\ . \label{abc}
\end{equation}
Modes with angular momentum on the $T^{1,1}$ directions also have
energy given by the same scale.  From our discussion in section
\ref{s:geometry}, we note that $ Rm_{f} \gtrsim 1 $, so the lowest
mass is about as large as the KK mass gap.  Thus, the lowest mass
mode of a massive bulk scalar (such as the dilaton-axion) must be
integrated out along with the higher KK modes for a large class of
applications.

\subsection{Semiclassical Localization}\label{s:semiclassical}

     Next, we would like to discuss the localization of modes in
more general settings. We can see when localization occurs in a
rough semiclassical approximation using the Schr\"odinger-like
equation (\ref{dilax-x}). Using the effective potential, we can
also discuss the effects of the variation of the flux induced mass
term in the equation of motion of the dilaton. Recall that the
potential is given by (\ref{dilax-V}) as
\be
\lb{dilax-V2} V(x) =
e^{-3\t A/2}\del_x^2 e^{3\t A/2}+e^{2\t A}\left(
m_f^2(x)+\frac{Q^2}{L^2}\right)\ ,
\ee
where $m_f(x)$
measures the magnitude of the flux in the warped geometry and $Q$
measures the angular momentum of the KK mode in the other 5
dimensions of the compact manifold. Most qualitative features of
the spectrum can be inferred  from the structure of $V(x)$.

   As an illustration, consider a pure $\mathnormal{AdS}_5$ throat,
as in the KT geometry.  The potential (\ref{adsp}) is inverse
square and has a global minimum at $ x_0 = R^{2}/r_{0}$, the
infrared end of the warped extra dimension.  We would then expect
the low lying modes to have masses at the scale set by the value
of the potential at this minimum. This indeed is the case as can
be seen from the results of our explicit computation, equations
(\ref{adsm}) and (\ref{abc}).  That is, the dilaton zero-mode has
a mass scale $m\sim e^{A_0}m_f = m_f R/x_0$.

 We are now ready to incorporate the effects of the variation
of the flux term. For the lowest mode the relevant terms in the
potential are the terms involving the warp factor and the flux
induced mass term. Note that  the potential $V(x)$ is a local
function of the warp factor and the fluxes. The parameters of our
$AdS \times T^{1,1}$  model describe the throat
region accurately; thus, even in the case of a realistic warp
factor and flux-induced mass terms, the potential in the throat
region is given by
\begin{equation}
   V_{\mathnormal{throat}}(x) = \frac{1}{x^2}\left(\frac{15}{4} + m_f^2R^2
 \right)\ .
   \label{adspt}
\end{equation}
  In the bulk region, the derivatives of the warp factor are
small, and the dominant contributions arises from the second term
in (\ref{dilax-V}). From our discussion of the bulk geometry and
fluxes in section \ref{s:geometry} it is easy to see that
\begin{equation}
      V_{\mathnormal{bulk}}(x) \approx  \frac{m'^{2}_{f}}{c^{1/2}}
= \frac{g_s n_f^{\prime 2}}{\ap c^2}\ .
\end{equation}
Comparing this with the value of (\ref{adspt}) at its minimum , we
conclude that the global minimum of the potential is in the
infrared end of the throat if $e^{A_{0}}  \ll 1/c$. On the other
hand, if $e^{A_{0}} \gg 1/c$, then the minimum of the potential is
in the bulk.

      To summarize, our models suggest that  the flux induced
mass for the dilaton is determined by an interplay between the
parameter $c$ (which controls the overall volume) and
$e^{A_{0}}$ (the minimum value of the warp factor). In order
understand this competition, it is useful to start in the infinite
volume limit $(c \gg e^{-4A_{0}})$ and decrease $c$ adiabatically.
In the infinite volume limit,  the dilaton acquires a mass
\begin{equation}
     m \approx   \frac{m'_{f}}{c^{1/4}}
=\sqrt{g_s}\frac{n^{\prime}_f}{\sqrt{\ap} c}\ , \label{bulkm}
\end{equation}
and its wavefunction in the internal directions is approximately
constant.   This is a roughly string scale mass (since $n'_f$ is not very
large), suppressed by a volume factor.

Now consider decreasing the parameter $c$. For $ c \sim
e^{-4A_{0}}$, the geometry begins to develop a throat region. As
$c$ decreases further, the throat becomes more and more
distinct. In this process, the wavefunction and mass of the
dilaton also vary continuously.  For $ c \ll e^{-A_{0}} $, a
qualitatively distinct behavior sets in . The dilaton-axion
wavefunction becomes highly localized, and the mass is
\begin{equation}
         m \approx m_{f} e^{A_{0}}=\sqrt{g_s}\frac{n_f}{\sqrt{\ap}}e^{A_0}\ .
\label{deepm}
\end{equation}
This mass is approximately the warped string scale.
The mass formulae  (\ref{bulkm}) and (\ref{deepm}) agree for $c
\sim e^{-A_{0}} $, suggesting a smooth interpolation between
the two regimes.  In fact, as we will see in section \ref{s:detailed}
below, this semiclassical reasoning can fail in the regime
\be\lb{quantum}
e^{-A_0}\lesssim c \lesssim e^{-4A_0}\ .\ee

  Finally, we would like to make some comments on mixing between zero-modes
and naive kk modes encountered in the
linearized analysis of \cite{hep-th/0507158}. At large volume,
the lowest mass wavefunctions are given by the zero-modes of the Calabi-Yau.
On the other hand, in the deep throat regime, the wavefunctions
are localized. A Fourier decomposition of such a function in
terms of the eigenmodes of the underlying Calabi-Yau $(\tg_{mn})$
will involve non-trivial contributions from kk modes. This picture
seems consistent with the strong mixing between zero-modes and kk
modes found in \cite{hep-th/0507158} while examining massive
fluctuations in the presence of deep throats. The effect of these
kk modes is to localize the wavefunctions in the region of large
warping, an energetically \textit{favorable} process.

\subsection{More Detailed Models}\label{s:detailed}

 In the discussion of section \ref{s:simpleeom},
we modeled the bulk by imposing
boundary conditions at $r=r_{c}$. However, we would be remiss if
we did not explicitly add a region of constant warp factor
representing the bulk. The
model then becomes similar to the toy example discussed in
the introduction.
\comment{We do not present the details here, but it is
easy to to check that the low energy spectrum is still described
by (\ref{adsm}) and (\ref{abc}).}

The metric is of the general form (\ref{gkpm}) with the simplification
(\ref{6dmetric}), as before, with an $\mathnormal{AdS}_5$ warp
factor $e^A=r/R$ dominating in a throat region and a constant
warp factor in the bulk region.  In summary,
\be\lb{boxwarp}
e^{\t A} \equiv \left[c+e^{-4A}\right]^{-1/4} =
\left\{ \begin{array}{ll} e^A = r/R& (r_0<r<r_c)  \\
c^{-1/4}& (r_c<r<r_c+\sqrt{\ap})\end{array}\right.\ .
\ee
For continuity of the warp factor, we demand
$r_c/R = c^{-1/4}$, as in section \ref{s:simpleeom}.  Also as before,
$e^{A_0} = r_0/R$ and $e^H=e^{-A_0}/c^{1/4}$.  The proper length of the
bulk region is $c^{1/4}\sqrt{\ap}$, consistent with the role of $c$ as
a radial modulus.  As we saw in section \ref{s:geometry}, the flux-induced
mass is constant in the throat but is suppressed by the warp factor in the
bulk:
\be\lb{mf}
m_f^2(r) = \left\{ \begin{array}{ll} \b m_f^2 & (r_0<r<r_c)\\
\b m_f^2/c^{3/2}& (r_c<r<r_c+\sqrt{\ap})\end{array}\right.\ ,\ee
where $\b m_f=\sqrt{g_s}n_f/\sqrt{\ap}$ is a constant mass.
This formula for the
mass represents the extreme possibility that the dilaton-axion feels
the same flux in the bulk and throat regions.

In terms of the Schr\"odinger equation (\ref{dilax-x}), we find
$x=R^2/r$ in the throat region and $x=2x_c -\sqrt{c}r$ in the bulk region,
with $x_c=R^2/r_c = c^{1/4}R$.  Signs and integration constants have been
chosen so that the coordinate $x$ is continuous as a function of $r$.
For states with no angular momentum, the potential is
\be
V(x) = \frac{1}{x^2}\left( \frac{15}{4}+\b m_f^2 R^2\right)\Theta(x-x_c)
+ \frac{\b m_f^2}{c^2}\Theta(x_c-x)
+\delta(x-x_c)
\left(\frac{x}{R}\right)^{3/2}\del_x\left(\frac{R}{x}\right)^{3/2}
\ .\lb{box-V}
\ee
The final term requires that
\be\lb{box-bc}
\psi(x)\ \textnormal{and}\ \del_x\left( e^{3\t A/2}\psi\right)(x)\ee
are continuous at the boundary point $x_c$.  These boundary conditions
are consistent with (\ref{bdry-r}) for a continuous warp factor.
We also impose Neumann boundary conditions on $\delta\tau = e^{-3\t A/2}\psi$
at the endpoints
$x_c-\sqrt{c \ap}$ and $x_0$.

\FIGURE[t]{\includegraphics[scale=1]{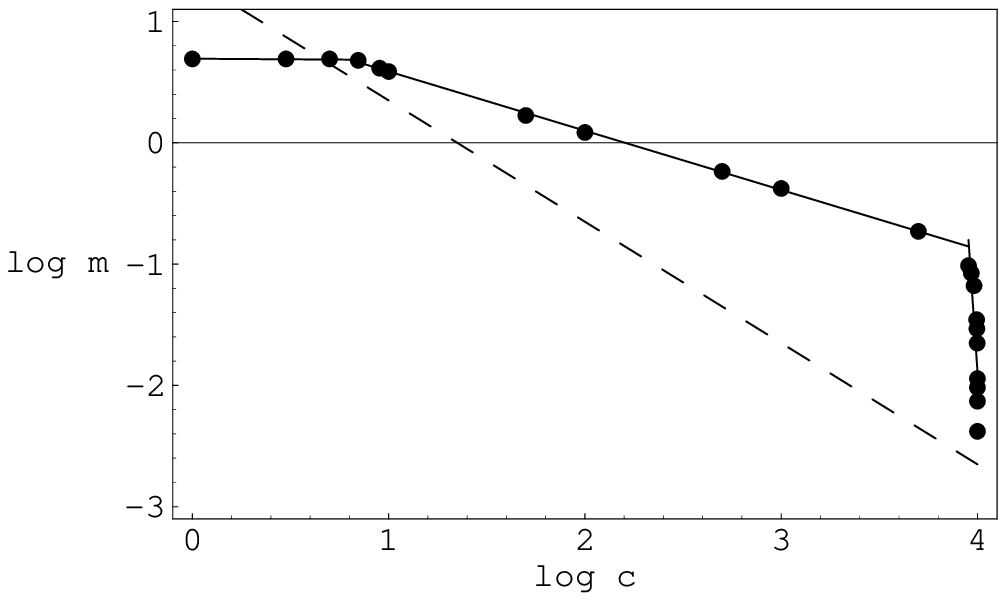}
\caption{\lb{f:box1}Log-log plot of mass versus volume modulus in
the regime $1 < c < e^{-4A_{0}}$. Mass is measured in units of the
warped KK scale $e^{A_0}/R$. Points represent the lightest masses
calculated in appendix \protect\ref{s:throatbulk}. The black
best-fit lines have slopes -0.01, -0.49, and -24 respectively, The
dashed line is the mass predicted by semiclassical bulk arguments
for $ c \gg e^{-A_{0}} $, $m=\b m_f/c$}}

We have calculated the lightest mass for this potential as a function
of bulk size $c$ for a fixed warp factor $e^{A_0}$ and flux-induced mass
$\b m_f$.  Since the throat
region disappears for $c\gtrsim e^{-4A_0}$ and the bulk should be at least
string scale, we consider only $1\leq c\leq e^{-4A_0}$.
We leave the details of the calculation for appendix \ref{s:throatbulk},
but we present the basic results here.

First, referring to figure \ref{f:box1}, we see that the lightest modes for
$c<e^{-A_0}$ do have mass $m\sim \b m_f e^{A_0}$, nearly independent of
the overall compactification volume $c$.  These modes are localized, but
it is easy to check that they delocalize when $c\gtrsim e^{-A_0}$, just
as we expect from the discussion of section \ref{s:semiclassical}.
However, the delocalized bulk modes have masses that scale like
$m\propto 1/\sqrt{c}$ rather than $m\propto 1/c$, so the exact results
do not follow our semiclassical intuition.  Then the masses drop sharply to
$m\sim \b m_f/c$ just before the throat disappears ($c\lesssim e^{-4A_0}$).
We should note that this interesting behavior is almost certainly
model-dependent, but the main point is that the intermediate behavior
indicates that the quantum mechanics of the potential (\ref{box-V}) are
important for some values of the compactification volume.

\FIGURE[t]{\includegraphics[scale=1]{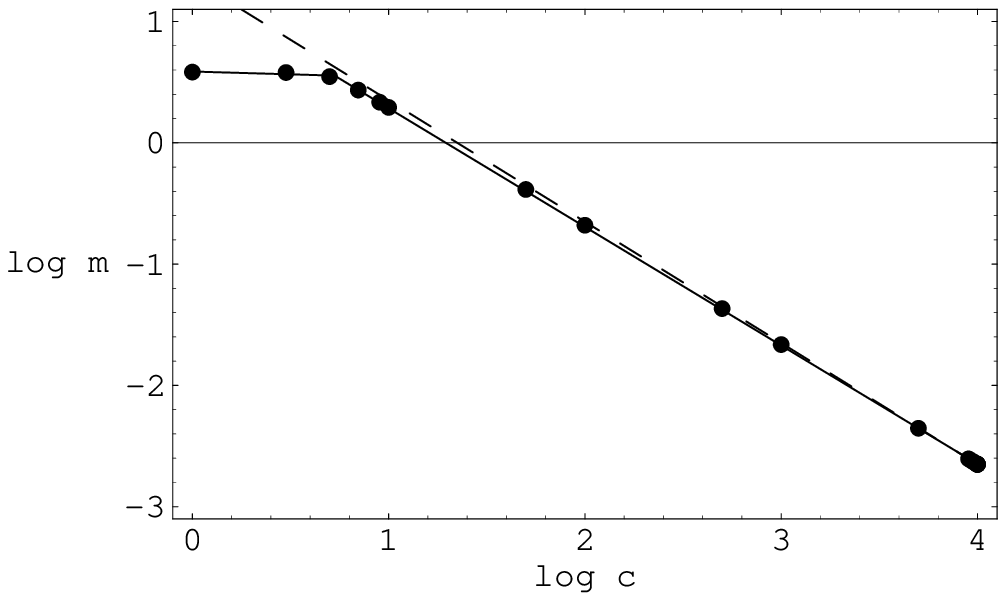}
\caption{\lb{f:box2}Log-log plot of mass versus volume modulus, as in
figure \protect\ref{f:box1}.  The best-fit lines have slopes -0.05 and
-0.98 respectively.}}

Because the flux described in the KS background cannot freeze the
dilaton by itself \cite{hep-th/0105097}, it is possible that the
flux-induced mass parameter differs in the bulk and the throat (or
perhaps decays as the warp factor decreases).  To check that our
qualitative results are not altered, we have also calculated the
lightest mode masses in the same geometry with \be\lb{mf2}
m_f^2(r) = \left\{ \begin{array}{ll} 0 & (r_0<r<r_c)\\
\b m_f^2/c^{3/2}& (r_c<r<r_c+\sqrt{\ap})\end{array}\right.\! .\ee
This is the extreme case that the dilaton-axion feels no flux in the
throat region.
In this case, we again find that the lightest mode masses
are constant with $c$ and approximately
$m\sim \b m_f e^{A_0}$ for $c\lesssim e^{-A_0}$, just as we expect from
semiclassical reasoning.  Then, for larger $c\gtrsim e^{-A_0}$, the lightest
modes have mass given by the semiclassical bulk result $m\sim \b m_f/c$.
See figure \ref{f:box2}.
Interestingly, the lightest modes never delocalize in the sense that they
always decay in the bulk; however, for $c\gtrsim e^{-A_0}$, the decay
is negligible, and the wavefuction is almost flat in the bulk.
Additionally, the KK wavefunction is much larger in the bulk than the throat,
so the modes have essentially delocalized.  See appendix \ref{s:throatbulk}
for a discussion of the wavefunctions.

The chief result of these calculations is that, by and large, semiclassical
physics is sufficient, although, in realistic models, quantum effects may
be relevant in the intermediate region $e^{-A_0}\lesssim c\lesssim e^{-4A_0}$.
Most importantly, the wavefunctions are localized in the throat region
for $c\lesssim e^{-A_0}$, where the mass is given by the warped scale,
$m\sim \b m_f e^{A_0}$.

In appendix \ref{s:throatbulk}, we also discuss the Kaluza-Klein mass gap
for the dilaton given a range of compactification parameters.  In case of
either flux-induced mass function (\ref{mf}) or (\ref{mf2}), we get similar
results for the KK mass gap for the lowest mass states:
\be\lb{kkgap1}
c\lesssim e^{-2A_0} \ \Rightarrow\ \Delta m_n \approx \frac{e^{A_0}}{R}\ee
and
\be\lb{kkgap2}
c\gtrsim e^{-2A_0}\ \Rightarrow\ \Delta m_n \approx \frac{1}{\sqrt{c\ap}}\ .
\ee
In the long throat ($c\lesssim e^{-A_0}$) regime, the gap (\ref{kkgap1})
is very important, since the lightest mode has a mass of the same
order as the KK mass gap.  In other words, it may be impossible to
describe moduli stabilization through an effective superpotential for
compactifications with long throats --- we must integrate out the stabilized
moduli at the same time as KK modes.  A similar concern arises
for the smaller KK mass gap (\ref{kkgap2}) if the lowest mass follows the
pattern of figure \ref{f:box1}, when the lightest mass will be of the same
order as the mass gap.  While our results only follow from clearly rough models
of the physics, they do caution that we may not be able to describe moduli
stabilization in effective theory when throats exist ($c\lesssim e^{-4A_0}$)
or are long ($c\lesssim e^{-A_0}$).  We also find in appendix
\ref{s:throatbulk} that it is possible to have, for large $m_f$, a
KK mass gap of $\Delta m\sim 1/\b m_f\ap$ at small $c$, which is suppressed
from the string scale by the flux quanta.  However, this can only occur
when $\b m_f\ap/R\gtrsim e^{-A_0}$.

\section{Discussion}\label{s:discuss}

In this section, we will review our findings and discuss their
implications for both the theory and phenomenology of flux
compactifications in string theory.  As a brief review, we recall
that we have found that the lowest dilaton-axion KK mode
wavefunction localizes in deep throats --- those for which
$e^{-A_0}\gg c$.  These localized modes have masses $m\sim
e^{A_0}/\sqrt{\ap}$.  For larger compactifications, $c\gtrsim
e^{-A_0}$, the mass decreases (see figures \ref{f:box1} and
\ref{f:box2}).  Semiclassically, we expect $m\sim 1 /c\sqrt{\ap}$,
which is the result in the infinite volume limit $c\gg e^{-4A_0}$, in
which the throat disappears. The semiclassical behavior may or may
not hold for $c\lesssim e^{-4A_0}$; in one of our examples, the
mass scales like $c^{-1/2}$, similar to a bulk kk mode. We have
also estimated the effect of compensators on the dilaton-axion
mass, and we found that they give a negligible contribution in the
bulk and sufficiently long throats. Our results were derived using
simple models, which nonetheless seem to encapsulate key features
of the corresponding string compactifications.

First, our results are important for determining which degrees of
freedom should be included in the four-dimensional effective
theory describing the compactification. The regime of validity of
the effective theory is given by the scale of the KK excitations
(since it describes the dynamics with these degrees of freedom
integrated out). In the deep throat regime, we found that there is
no hierarchial separation between the dilaton mass and its KK
scale. In fact, following \cite{hep-th/9906064,hep-th/0006191,
hep-th/0512076,hep-th/0512249,hep-th/0602136}, 
we expect that the KK excitations of the 4D graviton will also
have masses of this same scale. Therefore, we expect that all the
dilaton-axion should be integrated out of the effective theory in
the deep throat regime ($c\ll e^{-A_0}$).

On the other hand, the dilaton mass and KK scale separate at larger
compactification volume.  Following our semiclassical analysis, we
expect a dilaton mass $1/c\sqrt{\ap}$ in the short-throat and no-throat
regime.  This compares to a KK mass scale of $e^{A_0}/\sqrt{\ap}$
(for $c\lesssim e^{-2A_0}$) or $1/\sqrt{c\ap}$ (larger $c$).  In this case,
the dilaton-axion need not be integrated out of the effective theory.
We do caution that semiclassical reasoning may fail for
$e^{-A_0}\lesssim c\lesssim e^{-4A_0}$, in which case the two scales may not
be hierarchically separated.

So far, we have focused on the dilaton-axion. Widening our view a bit, the
equations describing the fluctuations of the metric moduli have also been
obtained in  \cite{hep-th/0507158}. These equations are quite
complicated, and we are unable analyze them in detail at present.
However, they are qualitatively similar to the equation for
the fluctuation of the dilaton-axion. In particular, the structure is
compatible with solutions which are oscillatory in a throat
region and damped in the bulk. This suggests that some (or all) of the
metric moduli might also exhibit the localization we have discussed
for the dilaton.

In fact, we expect very similar behavior for complex structure
moduli. In the infinite volume limit, the geometry is same as that
of the conventional Calabi-Yau construction. In this regime, the
equations for linearized fluctuations obtained in
\cite{hep-th/0507158} are easily analyzed, and the mass is just
$m_z\sim 1/c\sqrt{\ap}$. As the compactification volume is
decreased and a long throat develops, one expects the
wavefunctions to change; becoming highly localized in the throat
region. The masses would then redshift to $m_z\sim
e^{A_0}/\sqrt{\ap}$. The details will most likely depend on the
mode in question and are beyond the scope of our analysis, so it
is possible that some moduli will not fit this pattern, in which
case they are more massive than KK modes in the deep throat
regime. However, the fact that the dilaton --- whose wavefunction
is uniform in the infinite volume limit --- localizes indicates
that this behavior might be fairly generic.

The dynamics of moduli is crucial for understanding
physics related to the hidden sector. In the context of IIB
compactifications, it can be argued that the superpotential
generated by the fluxes is that of Gukov, Vafa, and Witten
\cite{hep-th/9906070},
\begin{equation}
   W \propto \int G \wedge \Omega\ .
\end{equation}
This superpotential describes the
effective field theory of the dilaton-axion and complex structure
moduli, with higher Kaluza-Klein modes integrated out.  We have argued that
if one starts at the infinite volume limit and continuously
decreases the volume, the wavefunctions of the the dilaton and
possibly some of the metric moduli continuously localize to the
bottom of the longest throat. Given the non-renormalization properties of
the superpotential, we expect it to be relevant for the dynamics of
modes localized in the deepest throat, rather than bulk modes or modes
localized in other throats.

While the superpotential is thought to be unaffected by the warp factor,
\comment{(see \cite{hep-th/0208123} for a direct calculation),}
our results show that warping corrects the K\"ahler potential.  Taking
the uncorrected K\"ahler potential of an orientifold Calabi-Yau 
compactification just gives a mass $m\sim n_f/c\sqrt{\ap}$ 
for the dilaton-axion.
However, as we have seen, the lowest dilaton 
mass is given by the warped string scale
$m\sim n_f e^{A_0}/\sqrt{\ap}$ when there are long throats, so the
K\"ahler potential must be corrected.  Since we have argued that
compensators should give only small contributions to the mass, 
a possible 
K\"ahler potential is that proposed in \cite{hep-th/0208123}, with suitable
modifications due to corrected holomorphic coordinates, as suggested in
\cite{hep-th/0507158}.  We should
caution, though, that the dilaton-axion (and complex structure moduli)
may be too heavy for the effective theory, in which case
only classically massless fields would appear in the K\"ahler potential.

\comment{
An intriguing possibility is that corrections to the potential might 
be better described as corrections to the superpotential rather than to
the K\"ahler potential\footnote{We thank J. Polchinski for conversations on 
this point.} (although the only physically meaningful quantity is
the combination $\mathcal{K}+\ln |W|^2$).  One reason we might think of
the superpotential as being corrected is that the minimum warp factor of
a throat is nonperturbatively small in the string coupling 
($e^{A_0}\sim e^{-B/g_s}$ for $B$ a ratio of flux quanta), so the warping
could induce a nonperturbative correction to $W$.  We can also give another
line of reasoning: usual arguments for the superpotential to be uncorrected
perturbatively rely on renormalization group flow.  In the case of
unwarped compactifications, rescaling the compactification volume is 
precisely a renormalization group flow because the KK masses scale with
the inverse radius, so the superpotential must be
corrected only nonperturbatively as the compact manifold shrinks.  However,
for a warped compactification, we have seen that KK masses do not 
continue to scale with the volume modulus when the warping is significant.
Thus, scaling the volume modulus is not a renormalization flow, and 
nonrenormalization theorems for the superpotential may not apply in a 
straightforward manner.  We take an agnostic point of view and assume that
the superpotential is not corrected by the warping, but the possibility 
should be mentioned.}

Our findings are also important in understanding the nature of
couplings between the standard model and the supergravity sectors in brane
world scenarios. Consider a scenario in which the standard model
branes are located at the bottom of a deep throat, with the
hierarchy problem solved as proposed by Randall and Sundrum.
In such a scenario, our calculations suggest that the mass of the
bulk modes would also be in the TeV scale, so we would expect
couplings of order unity between supergravity and matter
fields at the TeV scale. This is analogous to the findings of
Goldberger and Wise in \cite{hep-ph/9907218}.
The phenomenological implications
for a scenario in which the standard model branes are located in the
bulk of a compactification with a deep throat are also
interesting. In this case, the wavefunctions of the moduli are
highly suppressed in the bulk region, and their couplings to the
standard model branes would be very weak.  However, it would be
interesting to examine the collective effect on the of the large
number of localized modes on the bulk to brane coupling.  In terms of
string compactifications, D7-branes intersecting in the bulk could
provide the standard model degrees of freedom.

We can also consider cosmological consequences of these results.
For example, KKLT \cite{hep-th/0301240} models see common use in
brane-antibrane inflation scenarios (see, for example,
\cite{hep-th/0308055,hep-th/0105204,hep-th/0312020,hep-th/0403119}).  
The key element in
these models is an anti-D3-brane present at the bottom of a
KS-like throat, and the inflationary potential is given by the redshifted
brane tension.  In a high-scale inflation model, it is reasonable to 
expect a hierarchy of $e^{A_0}\sim 10^{-3}$ and $c^{1/4}\sim 10$, where
$c^{1/4}$ is also the linear scale of the bulk.  This clearly falls into 
the short throat regime we have discussed, in which the dilaton mass
is less than the warped KK scale.  
\comment{This fact may have interesting
consequences, since \cite{hep-th/0507257} have argued that
reheating might overpopulate stable graviton KK modes in the throat.  
However, including the dilaton and other frozen moduli as lighter degrees
of freedom in the analysis of reheating may change the picture somewhat.}
Reheating bulk modes would also provide a mechanism to reheat the standard
model if it were located on D7-branes wrapping a bulk cycle.  In terms
of inflation, there are also consequences for multi-throat models
discussed in \cite{hep-th/0412040,hep-th/0507257,hep-th/0508229}.
As \cite{hep-th/0508139} has argued, 
light degrees of freedom in a second throat could
shorten that throat, reducing its hierarchy.  Our results strengthen that
conclusion by arguing that frozen moduli do have redshifted masses in
long throats.  Also as discussed in \cite{hep-th/0508139}, the evolution of 
these light moduli could lead to an additional stage of reheating,
and it remains to be seen whether a moduli overproduction problem could
occur.

Let us also mention 
KKLT models \cite{hep-th/0301240} in the context of the cosmological 
constant.  In order to have a discretuum of states, 
the value of the four dimensional cosmological
constant is related to the redshift factor of a 
throat, so it is likely that a large class of such models contain
very long throats. In such a case, it seems likely that moduli and KK 
scale would be very light, meaning that the standard model would most
likely couple to highly excited KK modes.  It would be interesting to know
if this poses any phenomenological constraint.

We close with a brief comment on future directions.  An obvious extension
of our work is to more realistic models of the geometry.  For example,
\cite{hep-th/0602296} gives a better approximation of the KS geometry.  Another
important direction would be to extend our analysis to other complex
structure moduli, as in \cite{ouyang}.
Also, in KKLT models, the overall volume is stabilized by introducing
energy sources localized on certain submanifolds in the bulk of
the Calabi-Yau. It would be interesting to study the nature of the
wavefunction and the mass of the lowest mode after the
introduction of such sources.

\comment{
It will be interesting to develop understanding the gauge
theory dual of the
phenomenon we have discussed. Since the wavefunction of the modes
are localized at the bottom of the throat, one might hope that
their dynamics is can be captured by some features of the infrared
of a confining gauge theory.}

\acknowledgments
 The authors would like to thank
A. Basu, I. Bena, H. Elvang, S. Giddings, S. Gukov, N. Mann, P. Ouyang,
J. Polchinski, A. Sen, and A. Virmani
for useful discussions and communications. 
The work of ARF is supported by the John A. McCone postdoctoral 
fellowship at the California Institute of Technology and also partially
by the Department of Energy contract DE-FG03-92-1ER40701.
The work of AM is
supported by the Department of Energy under contract
DE-FG02-91ER40618.  AM would like to thank the Hebrew University
for hospitality during the 23rd Jerusalem Winter School.

\appendix

\section{Estimate of Compensator Effects}\label{s:compapp}

In this appendix, we detail our estimate of the effects of compensators
on the dilaton-axion mass and wavefunction.  Since the NSNS and RR sectors
decouple and the equations have the same form in both sectors, we work
only with the NSNS sector equations (\ref{coupledns}) (remember that
we are setting the angular momentum to zero on the internal 5 dimensions).
Following the
logic leading to the Schr\"odinger-like form (\ref{dilax-x}) of the
dilaton-axion equation of motion, we can rewrite (\ref{coupledns}) as
the eigenvalue equation
\be\lb{eigen1}
H\left[\begin{array}{c} \psi_\phi(x)\\ \psi_B(x)\end{array}\right]
= \left(H_0+\delta H\right)\left[\begin{array}{c} \psi_\phi(x)\\
\psi_B(x)\end{array}\right]
=m^2 \left[\begin{array}{c} \psi_\phi(x)\\ \psi_B(x)\end{array}\right]\ee
with ``Hamiltonian'' and perturbation
\bea
H_0 &=& \left[\begin{array}{cc} -\del_x^2+\frac{1}{x^2}\left(
\frac{15}{4}+m_f^2 R^2\right)&\\
&-\del_x^2 +\frac{15}{4}\frac{1}{x^2}\end{array}\right]\ ,
\nonumber\\
\delta H&=&\left[ \begin{array}{cc} &-\frac{1}{6}\frac{n_h}{R}
\left(\frac{R}{x}\right)^{5/2}\del_x\left(\frac{x}{R}\right)^{3/2}\cdot\\
2\frac{n_h}{R}\left(\frac{R}{x}\right)^{9/2}\del_x
\left(\frac{x}{R}\right)^{3/2}\cdot& \end{array}\right]\ .\lb{eigen2}
\eea
The $\cdot$ in the perturbation Hamiltonian indicates that the derivative
should act on the wavefunction as well as the explicit factor.

The other important contribution to the problem comes from the boundary
conditions.  We have Neumann boundary conditions for $\delta\phi$ at both
ends of the throat, meaning
\be\lb{neumann-ads}
\del_x\left(\left(\frac{x}{R}\right)^{3/2}\psi_\phi\right)(x_0)
=\del_x\left(\left(\frac{x}{R}\right)^{3/2}\psi_\phi\right)(R)=0\ .
\ee
The 2-form potential satisfies the same Neumann boundary conditions in
the IR end of the throat $x_0$, but we take it to satisfy Dirichlet
boundary conditions at the UV end $x=R$:
\be\lb{mixed-ads}
\del_x\left(\left(\frac{x}{R}\right)^{3/2}\psi_B\right)(x_0)=0\ ,\ \
\psi_B(R)=0\ .\ee
This Dirichlet boundary condition simulates the odd parity of $B_{\mu\nu}$
under the orientifold 3-plane or 7-plane projection around a point in the
bulk.  Using (\ref{bessel-ads}), we find the unperturbed eigenstates
\bea
\ket{\phi,n} &=& \left[\begin{array}{c}
\sqrt{\frac{x}{R}}\left(C_1 J_q(m_\phi x)+C_2 Y_q(m_\phi x)\right)\\
0\end{array}\right]\nonumber\\
\ket{B,n} &=& \left[\begin{array}{c} 0\\
\sqrt{\frac{x}{R}}\left(D_1 J_2(m_B x)+D_2 Y_2(m_B x)\right)
\end{array}\right]\ ,\lb{unperturbed}\eea
where $q$ is defined in (\ref{bessel-ads}) and $n$ indicates the excitation
number in the $x$ direction.  The coefficients are related to each other
by the $x=R$ boundary conditions:
\be\lb{coeffs}
\frac{C_2}{C_1} = -\frac{2J_q(m_\phi R)+m_\phi RJ_q'(m_\phi R)}{2Y_q(m_\phi R)
+m_\phi RY_q'(m_\phi R)}\ ,\ \
\frac{D_2}{D_1} = -\frac{J_2(m_B R)}{Y_2(m_B R)}\ .\ee
The masses are determined by transcendental equations obtained from
applying (\ref{coeffs}) to the $x=x_0$ boundary conditions
\cite{hep-ph/9907218}.

\FIGURE[t]{
\includegraphics[scale=1]{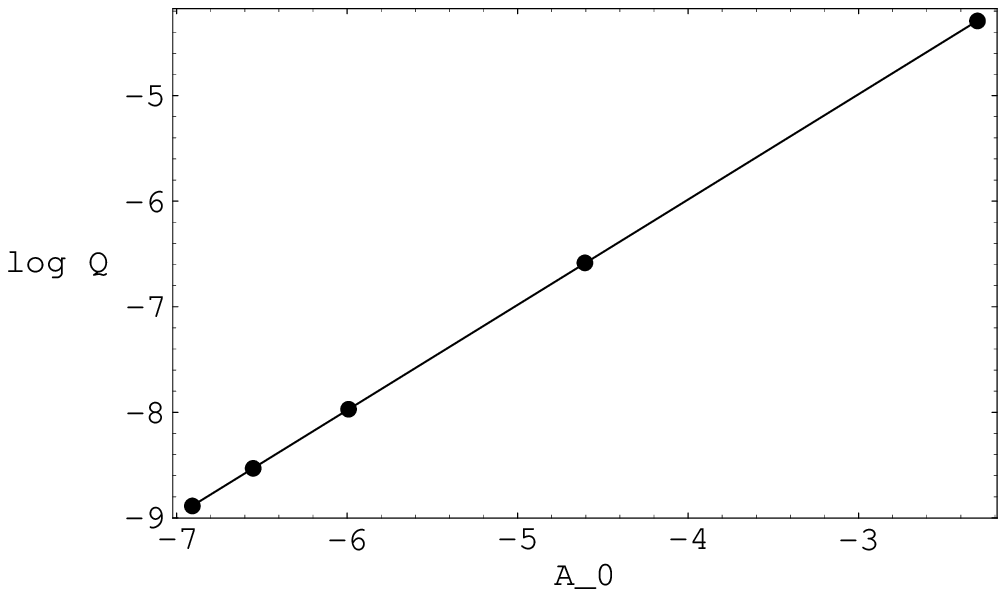}
\caption{\label{f:compensator} Change in mass-squared versus
minimal warp factor $A_0$.  $Q=|\Delta (m_\phi^2)/m_\phi^2|$.
Points are calculated values, and the line is the best-fit
$\ln Q = 0.998 A_0-1.99$.}
}

Despite the fact that the perturbation Hamiltonian is not hermitian, usual
perturbation theory still applies, as in quantum mechanics.  Normalizing
our wavefunctions to unity when integrated over $x$, we find
a second-order contribution to the lowest dilaton mass
\be\lb{perturb}
\Delta (m_\phi^2) = \sum_n \frac{\bra{\phi,0}\delta H\ket{B,n}
\bra{B,n}\delta H\ket{\phi,0}}{m_\phi^2 - m_{B,n}^2}\ .\ee
For a fixed hierarchy $e^{-A_0}=x_0/R=10^3$,
we have shown by direct calculation that the two lowest
modes of $\psi_B$ dominate the correction to the dilaton mass-squared.  Also,
given the behavior of the terms we have checked explicitly, the
sum appears to be well-behaved and convergent.  Throughout our calculations,
we have set $n_h=1$ and $m_f R=3$.

We now want to study the dependence of $\Delta (m_\phi^2)$ on the warped
hierarchy $e^{A_0}$.  To do so, we calculated the sum (\ref{perturb}) for
the two lowest modes of $\psi_B$ at five different values for the
minimum warp factor.  At these hierarchies, we find
\be\lb{ratio}
\left|\frac{\Delta (m_\phi^2)}{m_\phi^2}\right| \propto e^{A_0}\ .
\ee
This scaling is shown in figure \ref{f:compensator}.  This rather clean
scaling behavior is somewhat of a surprise, given the number of truncations
and assumptions we have made in arriving at our estimated $\Delta (m_\phi^2)$.
It is certainly an intriguing idea that the full effect of the compensator
might obey a simple scaling law with the warped hierarchy.  Finally,
given that the effect of the compensator is quite small (as can be seen
from the figure) even for relatively mild hierarchies, we will ignore
the effects of compensators throughout the rest of the paper.

\section{Throat Plus Bulk Calculations}\label{s:throatbulk}

In this appendix, we give specifics of calculations needed to understand
models with both throat and bulk regions, with the warp factor
given in equation (\ref{boxwarp}) and flux-induced mass parameter
given in (\ref{mf},\ref{mf2}).  We consider the case of vanishing
angular momentum ($Q=0$) for simplicity.

In the throat region, the wavefunction is given by Bessel functions
as in (\ref{bessel-ads}):
\be \psi(x) = \sqrt{\frac{x}{R}}J_{\nu}(m_n x)
 + b_{n} \sqrt{\frac{x}{R}}Y_{\nu}(m_n x)\ .\ee
Depending on the flux that couples to the dilaton in the
throat, we have
\be\nu = \sqrt{4 +\b m_f^2 R^2}\ \textnormal{or}\ \nu=2\ .\ee
By imposing Neumann boundary conditions at $x_0$ (the bottom of the throat),
we find
\be\lb{throat-Neumann}
b_n = -\frac{2J_\nu(m_nx_0)+m_n x_0J'_\nu(m_nx_0)}{2Y_\nu(m_nx_0)
+m_n x_0Y'_\nu(m_n x_0)}
\ .\ee

In the bulk region, imposing Neumann boundary conditions at $x_c-\sqrt{c\ap}$
gives two possible solutions
\be
\psi(x) = B_n \left\{ \begin{array}{ll} \cosh\left[\sqrt{\b m_f^2/c^2-m_n^2}
\left(x-x_c+\sqrt{c\ap}\right)\right]& (m_n\leq \b m_f/c)\\
\cos\left[\sqrt{m_n^2-\b m_f^2/c^2}
\left(x-x_c+\sqrt{c\ap}\right)\right]& (m_n\geq \b m_f/c)
\end{array}\right.\ .
\ee
The boundary conditions (\ref{box-bc}) at $x_c$ are therefore satisfied
when
\be\label{besseltan}
f(z_n)\equiv 2e^H +z_n \frac{(2Y_\nu(z_n)+z_nY'_\nu(z_n))J'_\nu(z_n e^{-H})
-(2J_\nu(z_n)+z_nJ'_\nu(z_n))Y'_\nu(z_n e^{-H})}{(2Y_\nu(z_n)+z_nY'_\nu(z_n))
J_\nu(z_n e^{-H})-(2J_\nu(z_n)+z_nJ'_\nu(z_n))Y_\nu(z_n e^{-H})}
\ee
equals
\be
\sqrt{\b m_f^2 x_0^2/c^2 -z_n^2}\tanh\left[ \sqrt{\b m_f^2 x_0^2/c^2 -z_n^2}
\frac{\sqrt{c\ap}}{x_0}\right]
\ee
or
\be\lb{ellipsetan}
-\sqrt{z_n^2-\b m_f^2 x_0^2/c^2}\tan\left[ \sqrt{z_n^2-\b m_f^2 x_0^2/c^2}
\frac{\sqrt{c\ap}}{x_0}\right]\ ,\ee
where $z_n=m_n x_0$.

We have found numerically the lightest mass modes for the fixed values
$\b m_f=\sqrt{5}$, $e^{A_0}=1/10$, and $\sqrt{\ap}/x_0=e^{A_0}$, and allowing
the size of the bulk $c$ (and therefore the hierarchy $e^H$) to vary
from $1$ to $e^{-4A_0}$.  We find the results presented in section
\ref{s:detailed}.  We also present here (figure \ref{f:waves}) the
wavefunctions of the lightest mass modes for several different values of
the volume modulus $c$.  In particular, note that the wavefunctions are
suppressed in the throat region ($x>x_c$) for $c\gtrsim e^{-A_0}$, but that
only the case in which $m_f(x)\neq 0$ in the throat yields an oscillatory
wavefunction in the bulk.

\FIGURE[t]{\includegraphics[scale=1.3]{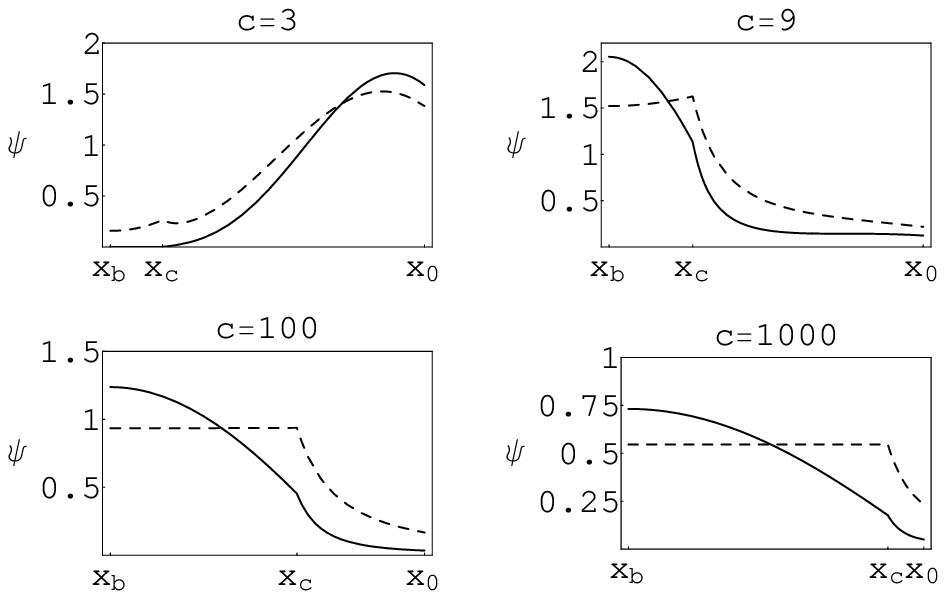}
\caption{\lb{f:waves}Wavefunctions for 4 values of $c$.  The solid curves
are for $m_f(x)\neq 0$ in the throat, while the dashed are for $m_f(x)=0$
in the throat.  For comparison, the wavefunctions have been normalized as
$\int dx |\psi|^2=1$.  In the figures, $x_b=x_c-\sqrt{c\ap}$.  Note that the
throat gets shorter and the bulk longer as $c$ increases.}}

Let us also discuss the mass spacing of KK modes in these two models.
A typical solution (note that the lowest solution $z_0$ may very well be
atypical) to the boundary condition (\ref{box-bc}) occurs near a singularity
of either $f(z)$ defined in (\ref{besseltan}) or the tangent
(\ref{ellipsetan}), so the KK mass gap is defined by the spacing of
singularities of those two functions.  The singularities of $f(z)$ are
spaced by $\Delta z\sim 1$ from our knowledge of Bessel functions, and
numerical work shows that the spacing is indeed around 4 or 5.
This spacing leads to a mass gap of order $\Delta m_n\sim 1/x_0 = e^{A_0}/R$,
which is the redshifted curvature scale for the throat.
On the other hand, the tangent (\ref{ellipsetan}) has singularities when
\be
\sqrt{z^2/x_0^2-\b m_f^2/c^2}\sqrt{c\ap} = \frac{p\pi}{2}\ ,\ \
p\in(2\mathbb{Z}+1)\ .\ee
Differentiating, we can approximate
\be
\Delta z \approx \frac{p\pi^2 x_0}{2c\ap} \left(\frac{p^2\pi^2}{4c\ap}
+\frac{\b m_f^2}{c^2}\right)^{-1/2}\ .\ee
There are two regimes for this solution:
\bea
\Delta z \approx \frac{\pi x_0}{\sqrt{c\ap}}&& (\ap \b m_f^2\lesssim
\frac{c\pi^2}{4})
\lb{spacing1} \\
\Delta z \approx \frac{\pi^2 x_0}{\b m_f\ap}&& (\ap \b m_f^2\gtrsim
\frac{c\pi^2}{4})\ .\lb{spacing2}
\eea
Since the smallest spacing of singularities determines the KK mass gap,
this last value (\ref{spacing2})
is only important for $\b m_f\ap\gtrsim x_0$, or
$\b m_f \ap/R\gtrsim e^{-A_0}$ (consider the comparison to the singularity
spacing of $f(x)$).  Because the throat curvature radius
$R$ is determined by the flux (as is $\b m_f$), we may expect
$\b m_f\ap/R\sim 1$,
although it is possible that (\ref{spacing2}) becomes relevant
for a Calabi-Yau with a large number of 3-cycles on which the flux can
wrap (and sufficiently small volume modulus $c$).  These two spacings give
KK mass gaps of
\be
\Delta m_n \approx \frac{1}{\sqrt{c\ap}}\ \textnormal{or}\
\Delta m_n \approx \frac{1}{\b m_f\ap}\ee
respectively.  These are the unwarped bulk KK scale and an almost string
scale mass --- it is suppressed by the amount of flux.

Comparing the singularity spacings given above,
the KK mass gap (without large $\b m_f$) is the warped KK scale of the throat
$\Delta m\sim e^{A_0}/R$ for small $c\lesssim e^{-2A_0}$ transitioning to
the bulk KK scale $\Delta m \sim 1/\sqrt{c\ap}$ if the compactification
radius is increased sufficiently. At large flux-induced mass $\b m_f$, the
KK mass gap is $\Delta m\sim 1/\b m_f\ap$ at small volume modulus and
$\Delta m \sim 1/\sqrt{c\ap}$ at large volume.

Finally, we should mention that it is straightforward to check that there are
no massless modes in the potential (\ref{box-V}) for $\b m_f\neq 0$.  The
unnormalized wavefunction in the throat would be
\be
\psi(x) = x^{1/2-\nu} + \frac{2-\nu}{2+\nu} x_0^{2\nu} x^{1/2+\nu}\ ,\ee
The coefficient on the second term comes from imposing Neumann boundary
conditions on $\delta\tau$ at $x_0$.  The wavefunction in the bulk region
is the hyperbolic cosine above with $m_n=0$.  The boundary conditions at
$x_c$ then imply
\be
\frac{\b m_f}{c}\tanh\left( \b m_f\sqrt{\ap/c}\right) = (4-\nu^2) \left(
\frac{1+(x_0/x_c)^{2\nu}}{(2-\nu)x_c+(2+\nu)(x_0/x_c)^{2\nu}x_c}\right)\ .\ee
This matching condition is inconsistent because $\nu\geq 2$
for nonzero $\b m_f$
(or angular momentum, for that matter).  The left-hand side is positive
definite,
and the right-hand side can be positive only when the denominator is
negative, or
\be
\frac{\nu+2}{\nu-2}\left(\frac{x_0}{x_c}\right)^{2\nu}<1\ ,\ee
which is impossible because $x_0>x_c$.

\bibliographystyle{h-physrev4}\bibliography{warpedmasses}

\begin{thebibliography}{10}

\bibitem{hep-th/0105097}
S.~B. Giddings, S.~Kachru and J.~Polchinski,
\newblock Phys. Rev. {\bf D66}, 106006 (2002), [hep-th/0105097].

\bibitem{hep-th/0201028}
S.~Kachru, M.~B. Schulz and S.~Trivedi,
\newblock JHEP {\bf 10}, 007 (2003), [hep-th/0201028].

\bibitem{hep-th/0201029}
A.~R. Frey and J.~Polchinski,
\newblock Phys. Rev. {\bf D65}, 126009 (2002), [hep-th/0201029].

\bibitem{hep-th/0301139}
P.~K. Tripathy and S.~P. Trivedi,
\newblock JHEP {\bf 03}, 028 (2003), [hep-th/0301139].

\bibitem{hep-th/0312104}
A.~Giryavets, S.~Kachru, P.~K. Tripathy and S.~P. Trivedi,
\newblock JHEP {\bf 04}, 003 (2004), [hep-th/0312104].

\bibitem{hep-th/0505260}
I.~Antoniadis, A.~Kumar and T.~Maillard,
\newblock hep-th/0505260.

\bibitem{hep-th/0506090}
D.~Lust, S.~Reffert, W.~Schulgin and S.~Stieberger,
\newblock hep-th/0506090.

\bibitem{hep-th/0308156}
A.~R. Frey,
\newblock hep-th/0308156.

\bibitem{hep-th/0405068}
E.~Silverstein,
\newblock hep-th/0405068.

\bibitem{hep-th/0509003}
M.~Gra\~{n}a,
\newblock Phys. Rept. {\bf 423}, 91 (2006), [hep-th/0509003].

\bibitem{hep-th/0307084}
S.~P. de~Alwis,
\newblock Phys. Rev. {\bf D68}, 126001 (2003), [hep-th/0307084].

\bibitem{hep-th/0312076}
A.~Buchel,
\newblock Phys. Rev. {\bf D69}, 106004 (2004), [hep-th/0312076].

\bibitem{hep-th/0407126}
S.~P. de~Alwis,
\newblock Phys. Lett. {\bf B603}, 230 (2004), [hep-th/0407126].

\bibitem{hep-th/0507158}
S.~B. Giddings and A.~Maharana,
\newblock hep-th/0507158.

\bibitem{hep-th/9906070}
S.~Gukov, C.~Vafa and E.~Witten,
\newblock Nucl. Phys. {\bf B584}, 69 (2000), [hep-th/9906070].

\bibitem{hep-ph/9907218}
W.~D. Goldberger and M.~B. Wise,
\newblock Phys. Rev. {\bf D60}, 107505 (1999), [hep-ph/9907218].

\bibitem{hep-ph/9905221}
L.~Randall and R.~Sundrum,
\newblock Phys. Rev. Lett. {\bf 83}, 3370 (1999), [hep-ph/9905221].

\bibitem{hep-th/0512076}
H.~Firouzjahi and S.-H.~H. Tye,
\newblock JHEP {\bf 01}, 136 (2006), [hep-th/0512076].

\bibitem{hep-th/0512249}
T.~Noguchi, M.~Yamaguchi and M.~Yamashita,
\newblock hep-th/0512249.

\bibitem{hep-th/0602136}
X.~Chen and S.-H.~H. Tye,
\newblock hep-th/0602136.

\bibitem{hep-th/0208123}
O.~DeWolfe and S.~B. Giddings,
\newblock Phys. Rev. {\bf D67}, 066008 (2003), [hep-th/0208123].

\bibitem{hep-th/0007191}
I.~R. Klebanov and M.~J. Strassler,
\newblock JHEP {\bf 08}, 052 (2000), [hep-th/0007191].

\bibitem{hep-th/0108101}
C.~P. Herzog, I.~R. Klebanov and P.~Ouyang,
\newblock hep-th/0108101.

\bibitem{hep-th/0205100}
C.~P. Herzog, I.~R. Klebanov and P.~Ouyang,
\newblock hep-th/0205100.

\bibitem{hep-th/0002159}
I.~R. Klebanov and A.~A. Tseytlin,
\newblock Nucl. Phys. {\bf B578}, 123 (2000), [hep-th/0002159].

\bibitem{hep-th/0502113}
S.~Franco, A.~Hanany and A.~M. Uranga,
\newblock JHEP {\bf 09}, 028 (2005), [hep-th/0502113].

\bibitem{hep-th/0210254}
A.~R. Frey and A.~Mazumdar,
\newblock Phys. Rev. {\bf D67}, 046006 (2003), [hep-th/0210254].

\bibitem{hep-th/0001033}
C.~Csaki, J.~Erlich, T.~J. Hollowood and Y.~Shirman,
\newblock Nucl. Phys. {\bf B581}, 309 (2000), [hep-th/0001033].

\bibitem{hep-th/0006191}
J.~D. Lykken, R.~C. Myers and J.~Wang,
\newblock JHEP {\bf 09}, 009 (2000), [hep-th/0006191].

\bibitem{hep-th/9906064}
L.~Randall and R.~Sundrum,
\newblock Phys. Rev. Lett. {\bf 83}, 4690 (1999), [hep-th/9906064].

\bibitem{hep-th/0301240}
S.~Kachru, R.~Kallosh, A.~Linde and S.~P. Trivedi,
\newblock Phys. Rev. {\bf D68}, 046005 (2003), [hep-th/0301240].

\bibitem{hep-th/0308055}
S.~Kachru {\em et~al.},
\newblock JCAP {\bf 0310}, 013 (2003), [hep-th/0308055].

\bibitem{hep-th/0105204}
C.~P. Burgess {\em et~al.},
\newblock JHEP {\bf 07}, 047 (2001), [hep-th/0105204].

\bibitem{hep-th/0312020}
H.~Firouzjahi and S.-H.~H. Tye,
\newblock Phys. Lett. {\bf B584}, 147 (2004), [hep-th/0312020].

\bibitem{hep-th/0403119}
C.~P. Burgess, J.~M. Cline, H.~Stoica and F.~Quevedo,
\newblock JHEP {\bf 09}, 033 (2004), [hep-th/0403119].

\bibitem{hep-th/0412040}
N.~Barnaby, C.~P. Burgess and J.~M. Cline,
\newblock JCAP {\bf 0504}, 007 (2005), [hep-th/0412040].

\bibitem{hep-th/0507257}
L.~Kofman and P.~Yi,
\newblock Phys. Rev. {\bf D72}, 106001 (2005), [hep-th/0507257].

\bibitem{hep-th/0508229}
D.~Chialva, G.~Shiu and B.~Underwood,
\newblock JHEP {\bf 01}, 014 (2006), [hep-th/0508229].

\bibitem{hep-th/0508139}
A.~R. Frey, A.~Mazumdar and R.~Myers,
\newblock Phys. Rev. {\bf D73}, 026003 (2006), [hep-th/0508139].

\bibitem{hep-th/0602296}
P.~Langfelder,
\newblock hep-th/0602296.

\bibitem{ouyang}
S.~B. Giddings and P.~Ouyang,
\newblock work in progress.

\end{thebibliography}
\end{document}